\documentclass[preprintnumbers]{article}

\usepackage[english]{babel}

\usepackage[a4paper,top=2cm,bottom=2cm,left=3cm,right=3cm,marginparwidth=1.75cm]{geometry}

\usepackage{amsmath, amssymb}
\usepackage{graphicx}
\usepackage{caption}
\usepackage{subcaption}
\usepackage{float}
\usepackage{jheppub, hyperref, cleveref}

\usepackage{color}

\newcommand{\diff}{\mathrm{d}}

\title{
Probing primordial black holes at high redshift with future gravitational wave detectors}

\abstract{
We analyze the detection prospects for potential Primordial Black Hole Binary (PBHB) populations 
buried in the Stellar-Origin Black Hole Binary (SOBHB) population inferred by the LVK collaboration. We consider different PBHB population scenarios and several future Gravitational Wave (GW) detectors. To separate the PBHB component from the SOBHB one, we exploit the prediction that the PBHB merger rate does not decline as fast as the SOBHB one at high redshift. However, only a tiny fraction of PBHB events may be resolved individually, and the sub-threshold events may yield an undetectable Stochastic GW Background (SGWB). For this reason, we determine the statistical significance of the PBHB contributions in the number of resolvable events seen in future Earth-based detectors and the SGWB measured at LISA. We find that the synergy between these probes will consistently help assess whether or not a sizeable PBHB population is present.}

\author[a]{Paolo Marcoccia,}
\emailAdd{paolo.marcoccia@uis.no}

\author[a]{Germano Nardini,}
\emailAdd{germano.nardini@uis.no}

\author[b]{Mauro Pieroni}
\emailAdd{mauro.pieroni@cern.ch}

\affiliation[a]{Department of Mathematics and Physics, University of Stavanger, NO-4036 Stavanger, Norway}
\affiliation[b]{Department of Theoretical Physics, CERN, 1211 Geneva 23, Switzerland}

\preprint{CERN-TH-2023-211}

\begin{document}

\maketitle

\section{Introduction}
\label{sec:Int}
\noindent LIGO's first Gravitational Wave (GW) detection labeled GW150914~\cite{Abbott:2016blz, LIGOScientific:2016emj} opened the gates to the world of GW astronomy. Since that detection, the sensitivity of LIGO detectors has increased considerably, and the Virgo~\cite{Virgo:2022fxr} and KAGRA~\cite{KAGRA:2020cvd} experiments have joined the network. At present, the LIGO-Virgo-KAGRA (LVK) network has identified over ninety  GW events involving Black Holes (BHs) and Neutron Stars (NSs)~\cite{LIGOScientific:2021djp} and has started constraining the statistical properties of the Stellar Mass Black Hole Binaries (SMBHB) population, although some bounds remain loose. Primarily, these limitations arise from the current detector sensitivity, which has allowed us to measure only a few events with high precision. The determination of the merger rate distribution at high redshift ($z \gtrsim 1$) is a striking example of these limitations. As a matter of fact, the properties of the population at $z \gtrsim 1$ are, currently, only guessed by using phenomenological models following the Star Formation Rate (SFR)~\cite{Madau:2014bja, Madau:2016jbv, Mangiagli:2019sxg}, leaving open space for the presence of both long time delays in the BHB formation~\cite{Fishbach:2021mhp} and a variety of populations with different redshift behaviours~\cite{Franciolini:2021tla, Hutsi:2020sol, Antonelli:2023gpu} (see also refs.~\cite{Cusin:2019jhg, KAGRA:2021duu, KAGRA:2021kbb, PhysRevD.103.043002, Babak:2023lro}). While this might be a reasonable assumption if all the events observed by current detectors are BHBs of Stellar Origin (SOBHBs), at least in principle, different scenarios are possible, leading to different high redshift behaviors. \\   

An intriguing alternative for BH formation is the possibility for BHs to form due to some cosmological processes occurring in the early Universe. These objects are typically dubbed Primordial BHs (PBHs) to differentiate from BHs produced in some late-time astrophysical processes. PBHs might form when strong scalar perturbations, \emph{e.g.}, generated by some inflationary model violating slow-roll, re-enter the Universe horizon, leading to the collapse of some regions of space~\cite{1967SvA....10..602Z, Hawking:1971ei, Carr:1974nx, Carr:1975qj}. Such a mechanism would make PBHs, and, in particular, PBH Binaries (PBHBs), totally independent of star formation processes. Different inflationary set-ups and early-universe histories can result, \emph{e.g.}, in vastly different PBH abundances and mass distributions (see, \emph{e.g.},~refs.~\cite{Khlopov:2008qy, Carr:2009jm,
Garcia-Bellido_2017,
Sasaki:2018dmp, 
Carr:2020gox, LISACosmologyWorkingGroup:2023njw} for reviews of PBH formation and constraints). 
Interestingly, depending on the formation scenario, PBHs can account for a substantial portion of the observed Dark Matter (DM) abundance \cite{Carr:2020xqk, Villanueva-Domingo:2021spv, Carr:2020gox, LISACosmologyWorkingGroup:2023njw}. Furthermore, at least one of the two BHs in some SMBHBs might be of primordial origin, and, more in general, a PBHB population might contribute\footnote{Notice, however, that LVK observations put tight constraints on the fraction of DM in PBHs for BHs in the stellar mass range~\cite{Ali-Haimoud:2017rtz,
Franciolini:2021tla, Franciolini:2022tfm}.} to the events currently observed by LVK detector~\cite{Bird:2016dcv, Clesse:2016vqa, Ali-Haimoud:2017rtz, Raidal:2018bbj, Hall:2020daa}. In such a case, some statistical properties and observable signatures might radically differ from those arising when all SMBHBs are SOBHBs. \\

Despite their radically different origin and phenomenology, PBHs are elusive at current GW detectors. 
The challenge is partly rooted in the lack of unquestionable, discriminating predictions at low redshift (see ref.~\cite{Franciolini:2021xbq} for a systematic procedure to assess the origin of BHBs). The predictions on PBHs and SOBHBs at low redshift are, indeed, loose due to the plethora of viable inflationary mechanisms and the numerous unknowns on the stellar-origin formation channels. On the contrary, a robust model-independent discrimination criterion exists at high redshift: at distances beyond the SFR peak ($z\simeq 2$), the SOBHB merger rate must fast decline, whereas the PBHB merger rate can keep growing~\cite{Ali-Haimoud:2017rtz, Raidal:2018bbj, Atal:2022zux, Young:2019gfc, Hutsi:2020sol, Bavera:2021wmw}. Remarkably, while the SFR peak is beyond the reach of the present LVK interferometers, it will be in the range\footnote{It is worth stressing, however, that while detecting high-redshift events will be possible with future detectors, accurately inferring their distances will not be trivial~\cite{Mancarella:2023ehn}.} of future Earth-based detectors~\cite{aplusdesign, Koushiappas:2017kqm, DeLuca:2021hde, Maggiore:2019uih, Ng:2020qpk, Ng:2022vbz, Branchesi:2023mws, Franciolini:2023opt, Martinelli:2022elq}. In addition, the Laser Interferometer Space Antenna (LISA)~\cite{LISA:2017pwj} might look for imprints of PBHs in the milliHertz band from individual events~\cite{Guo:2017njn, LISACosmologyWorkingGroup:2022jok, LISACosmologyWorkingGroup:2023njw} and the Stochastic GW Background (SGWB)~\cite{Garcia-Bellido:2017aan, Cai:2018dig, Bartolo:2018evs, Unal:2018yaa, LISACosmologyWorkingGroup:2023njw}. Moreover, LISA will be sensitive to the SGWB due to the incoherent superposition of the weak signals from the SMBHB population, which, including the contribution of binaries at high-redshift, brings information on the behavior of the population above the SFR peak~\cite{ Chen:2018rzo, Cusin:2019jhg, Lewicki:2021kmu, 
Bavera:2021wmw,
Perigois:2020ymr, Babak:2023lro}. \\

In the present paper, we discuss the detection prospects for the PBHB population using its high-redshift behavior~\cite{Ng:2021sqn, Ng:2022vbz, Ng:2022agi, Martinelli:2022elq}. Specifically, we study how future detectors might be able to identify PBHB populations beyond a certain \emph{Fiducial Population} of SOBHB (based on ref.~\cite{Babak:2023lro}), which is broadly compatible with the current LVK observations~\cite{LIGOScientific:2021djp, KAGRA:2021duu}. For this purpose, we consider some well-established PBHB population models~\cite{Atal:2022zux, Bavera:2021wmw} and show how different detectors will complementarily probe their parameter spaces. We expect our qualitative results to be independent of our fiducial population and PBHB models. However, the quantitative outcomes are contingent upon our population selections, so it will be worth repeating our analysis when the statistical properties of the SMBHB populations become less uncertain than they are today.\\ 

The paper is structured as follows: In~\cref{sec:PBHSubPopMod}, we describe the PBHB (sub)population models, and in~\cref{sec:Meth}, we describe the methodology that we adopt. Our results are presented in~\cref{sec:Results}, where we discuss the detectability of different PBHB populations with future Earth- and space-borne GW detectors. In particular, we demonstrate that the SGWB detection will give important complementary information on the presence of deviations from the fiducial model. We devote \cref{sec:Conclusions} to our final remarks and conclusions. Some technical details relevant to our analysis are described in Appendixes~\ref{app:FidMod}, \ref{app:ConvMaps}, \ref{sec:detectors}, and \ref{app:analytical_SGWB}.

\section{Population models}
\label{sec:PBHSubPopMod}

In this section, we present the population models that we analyze in this work. Following refs.~\cite{KAGRA:2021duu, Babak:2023lro}, for a given population, the number of expected sources in a given interval of redshift and parameter space is given by
\begin{equation}
    \frac{\diff^2N(z, \theta, \xi)}{\diff\theta \diff z } = R(z)\left[\frac{\diff V_c}{\diff z}(z) \right] \frac{\mathrm{T}^{\rm Det}_{\rm Obs}}{1 + z} \, p(\theta | \xi )\, ,
\label{eq:GeneralSourceGenerator}
\end{equation}
where $\rm T^{Det}_{\rm Obs}$ is the detector observation time, $\diff V_c / \diff z (z)$ is the differential comoving volume, $R(z)$ is the merger rate, and $p(\theta | \xi)$ is the Probability Distribution Function (PDF) for the source to have some specific values for the binary parameters (collectively denoted with $\theta$) given some population hyperparameters (collectively denoted with $\xi$)\footnote{Consistently with the most recent LVK analyses~\cite{LIGOScientific:2020kqk, KAGRA:2021duu}, we are assuming the PDF to be separable. In particular, we assume all other parameters not to depend on the redshift.}. Notice that since $p(\theta |\xi )$ is normalized, the number of events in a redshift interval $[z_{m}, z_{M}]$ is given by
\begin{equation}
    \Delta N_{z_{m}, z_{M}} = \int_{z_{\rm m} }^{z_{\rm M}}
    \frac{\diff N(z)}{\diff z } \, \diff z = \mathrm{T}^{\rm Det}_{\rm Obs} \int_{z_{\rm m} }^{z_{\rm M}} \, \frac{R(z)}{1 + z}\, \left[\frac{\diff V_c}{\diff z}(z) \right] \, .
    \label{eq:delta_N}
\end{equation}

More in detail, the PDF term $p(\theta | \xi )$ can be expressed as
\begin{equation}
    p(\xi | \theta) = p(m_1,m_2|\xi_{\rm Mass})  \times p(\theta_{\rm Angles}|\xi_{\rm Angles})  \times p(\theta_{\rm Spins}|\xi_{\rm Spins}) \,,
    \label{eq:GeneralGeneratorPDFs}
\end{equation}
where $m_1$ and $m_2$ are the two masses, $p(m_1,m_2|\xi_{\rm Mass}) $ is the mass function depending on some hyperparameters $\xi_{\rm Mass}$, $p(\theta_{\rm Angles}|\xi_{\rm Angles})$ is the angle PDF, and $p(\theta_{\rm Spins}|\xi_{\rm Spins})$ is the spin PDF. The sources are assumed to be isotropic in the sky, with inclination and polarization uniformly distributed in their considered prior. Eccentricity in the orbit is neglected throughout this work. The spin PDF is further expanded as
\begin{equation}
    p(\theta_{\rm Spins}|\xi_{\rm Spins}) = p(a_1, a_2|\xi_{\rm Spin \, Amplitude})  \times p(\cos(t_1),\cos(t_2)|\xi_{\rm Spin \, Tilt}) \; ,
\end{equation}
where $p(a_1, a_2|\xi_{\rm Spin \, Amplitude}) $ and $p(\cos(t_1),\cos(t_2)|\xi_{\rm Spin \, Tilt})$ are the spin amplitude and spin tilt PDFs, with $a_i$ and  $t_i$ (with $i=1,2$) denoting the normalized spin amplitude and the angle between the binary angular momentum and the spin of the body $i$, respectively. \\

In our analysis, the SMBHB population consists of the sum of the SOBHB and PBHB populations. We model each population using the framework outlined in \cref{eq:GeneralSourceGenerator} and neglect binaries with mixed origins. For the SOBHB population, we consider a fiducial population model, utilizing PDFs derived from the most recent LVK studies. We set their hyperparameters to the best-fit values as determined by these studies~\cite{KAGRA:2021duu}.
We remind that the LVK data, which we refer to as GWTC-3, provide no direct constraint at $z \gtrsim 1$. To extend our analysis to the redshift range $1 \lesssim z \lesssim 10$, we assume that the SOBHB merger rate $R_{\rm SOBHB}$ tracks the SFR and that the PDFs remain redshift-independent. Specifically, we employ the Madau-Dickinson merger rate~\cite{Madau:2014bja} with negligible time delay (see ~\cite{Dominik:2012kk, neijssel2019effect, Mapelli:2017hqk}) as our SFR.
\Cref{app:FidMod} provides further details about our approach.\\

Due to their early-universe origin, PBHBs have different statistical properties than SOBHBs, and, in particular, the PBHB merger rate is not expected to track the SFR. While several possibilities exist in the literature (see,\emph{e.g.}, ref.~\cite{Atal:2022zux}), for the PBHB merger rate we adopt~\cite{Raidal:2018bbj, Young:2019gfc}
\begin{equation}
R_{\rm PBHB}(z) = \varepsilon R_0 \left[ \frac{t(z)}{t(z = 0)}\right]^{-34/37} \;,
\label{eq:PBH_Rtz}
\end{equation}  
which corresponds to a power law in cosmic time $t(z)$~\cite{Hutsi:2020sol, Bavera:2021wmw}, normalized so that in $z=0$ it is $\varepsilon$-times smaller than the fiducial SOBHB merger rate $R_0$ at the same redshift. 
Alternatively, one can parametrize the rate by relating it to $f_{\rm PBH} \equiv  \Omega_{\rm PBH} / \Omega_{\rm DM}$, that is, the total PBH energy density over the DM energy density; see~\Cref{app:ConvMaps} for details.\\

\begin{figure}[t]
  \centering\includegraphics[width=1.\linewidth]{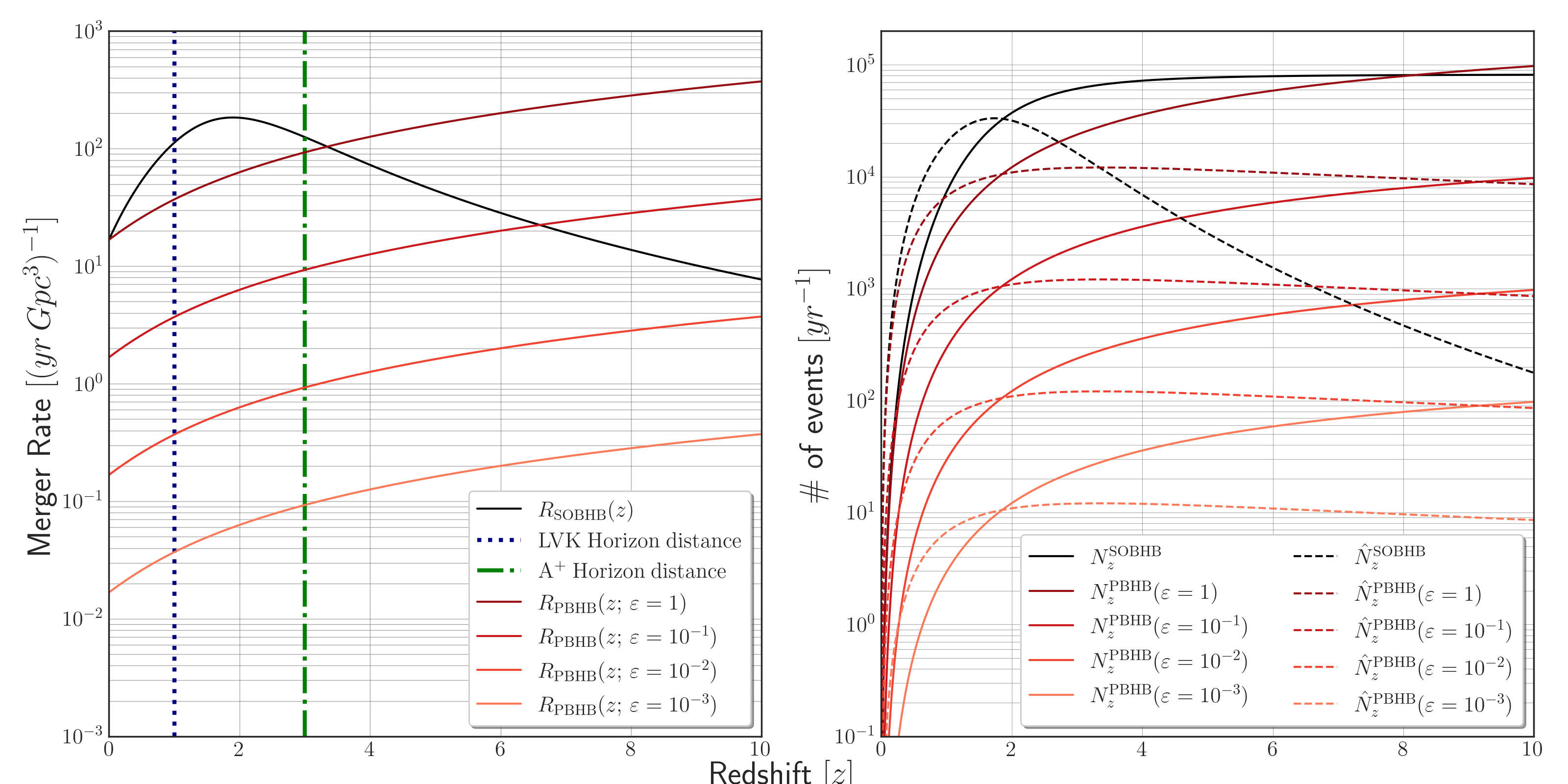} 
\caption{\small \emph{Left panel:} Merger rate as a function of redshift for different populations (solid curves). The dashed, vertical lines mark the maximal horizon distances of LVK (dotted blue line) and LIGO A${}^{+}$ (dash-dotted green line). \emph{Right panel:} number of events per year within a redshift volume (solid lines) and number of events per year per redshift bin normalized by the bin width (dashed lines)  as functions of redshift for different populations. In both panels,
all quantities relative to the fiducial SOBHB population are marked in black, and those relative to the PBH population with varying $\varepsilon$ are marked in red (dark to light for decreasing $\varepsilon$). }
\label{fig:RzComp}
\end{figure}

\Cref{fig:RzComp} shows the fiducial SOBHB merger rate and the PBHB merger rate for some values of $\varepsilon$. As long as $\varepsilon \ll 1$, the PBHB population is subdominant within the LVK horizon ($z\lesssim 1$). Thus, (up to a few outliers) the whole SMBHB population of GWTC-3 exhibits the SOBHB properties. On the other hand, the PBHB population may become relevant after the peak of the SFR. This is particularly evident from the dashed lines in the right panel of~\Cref{fig:RzComp}, which shows $\hat{N}^{\rm Pop}_{z}$ number of events redshift $z$ predicted by population Pop, which is given by
\begin{equation}
    \hat{N}^{\rm Pop}_{z} \equiv \frac{\Delta N^{\rm Pop}_{z_{m}, z_{M}} }{\Delta z} \, ,
    \label{eq:N_events_binned}
\end{equation}
where $\Delta N^{\rm Pop}_{z_{m}, z_{M}}$ is defined in~\cref{eq:delta_N} with the additional superscript specifying the considered population. By normalizing with $\Delta z=z_M-z_m$, we ensure that, for sufficiently small bin sizes, $\hat{N}^{\rm Pop}_{z}$ is independent of the specific binning scheme used in the analysis. \\

For the PBHB mass distribution, we assume each of the two masses in the binary to be drawn from the same PDF\footnote{This simplifying assumption doesn't include the effect of suppression terms in the mass function~\cite{Raidal:2018bbj, Young:2019gfc, Vaskonen:2019jpv, Franciolini:2022tfm, Franciolini:2023opt}. In the formalism of this paper, these effects are hidden in the definition of $\varepsilon$. For details see~\Cref{app:ConvMaps}.}, \emph{i.e.}, $P(m_1) = P(m_2)$, which we set to be a \emph{Log-Normal (LN)}
\begin{equation}
\Phi_{\rm LN} (m) = \frac{1}{\sqrt{2 \pi m^2 \sigma_{\rm LN}^{2}} } \exp\left[ -\frac{\ln^{2}(m/\mu_{\rm LN})}{2 \sigma_{\rm LN}^{2}} \right] \;,
\label{eq:PBH_LNMassPDF}
\end{equation}  
where $\mu_{\rm LN}$ and $\sigma_{\rm LN}$ are the hyperparameters setting the position and width of the peak. \\

Finally, we comment on the choices for the PDFs of the other variables. Since the PBHB spin PDF is still an argument of debate~\cite{DeLuca:2019buf, PhysRevD.104.063008, Saito:2023fpt},  we assume the PBHB spin distribution to be the same as the SOBHB one (see~\Cref{app:FidMod} and, in particular,~\cref{tab:spin_params}). Similarly, we assume all the angular variables (including the phase of the binary) to follow the same distribution of SOBHBs. 

\section{Methodology}
\label{sec:Meth}

This section describes our methodology to identify a PBHB population component on top of the fiducial population consistent with GWTC-3. Our analysis studies the detectability of individual sources and SGWB with future GW detectors. Specifically, we consider LIGO $\rm A^+$ and ET, with either $1$ or $10$ years of observations, for the measurement of individual sources, and LISA, assuming either $4$ or $10$ years of observations, for the SGWB detection and characterization. While our analysis concerns the PBHB population models presented in~\cref{sec:PBHSubPopMod}, a similar methodology could be applied to other PBH scenarios and GW detectors. For details on the detector characteristics and our numerical codes, see~\cref{sec:detectors} and the repository~\cite{GithubCodesPaolo}, respectively.

\subsection{Resolvable sources analysis}
\label{sec:IndSourc}

Our analysis of individually resolvable sources relies on measurements performed with Earth-based detectors. In particular, we check whether the presence of the PBHB population increases the expected number of detectable sources by more than $3\, \sigma$ beyond the number predicted by the fiducial population. For this purpose, hereafter we define an event as ``detectable'' when its Signal-to-Noise Ratio (SNR) (see definition in~\cref{eq:SNR}) is larger than some threshold value ${\rm SNR}_{\rm Thr}$, which we set to 8. \\

Let us start by briefly reviewing the methods to evaluate the SNR associated with a GW signal. The signal, $h$, measured by any GW detector, is expressed as
\begin{equation}
    h = F^{+}_{ij} h^+_{ij} + F^{\times}_{ij} h^\times _{ij} \;,
    \label{eq:DetSignal}
\end{equation}
where $h^{+}_{ij}$ and $h^{\times}_{ij}$ denote the two GW polarization modes, while  $F^{+}_{ij}$ and $ F^{\times}_{ij}$ are the detector pattern functions (for details, see \emph{e.g.}, ref.~\cite{Maggiore:1999vm}). The two GW modes can be further expanded as a combination of polarization tensors $\textrm{e}^{+}_{ij}$, $\textrm{e}^{\times}_{ij}$, and a waveform depending on the source parameters. In our study, we employ the IMRPhenomXHM waveform~\cite{Pratten:2020ceb}, a phenomenological waveform from the IMRPhenom family~\cite{Ajith:2007kx, Ajith:2007qp, Santamaria:2010yb, Khan:2015jqa, Husa:2015iqa}, offering a good compromise between quality and computation speed. With this choice, the signal depends on the parameters
\begin{equation}
\label{eq:signal_parameters}
    m_1, m_2, d_L, \phi_0, \tau_c, \theta, \phi, \iota, \psi,   \chi_1, \chi_2 \; , 
\end{equation}
where $m_1$ and $m_2$ are the masses of the two BHs in the detector frame, $d_L$ is the luminosity distance, $\phi_0$ is the binary's initial phase, $\tau_c$ is the coalescence time, $\theta$ and $\phi $ are the binary's latitude and longitude, $ \iota$ is the angle between the binary's angular momentum and the line of sight, $ \psi$ is the orientation, and $\chi_1$ and $\chi_2$ are the two (dimensionless) spin amplitudes projected on the orbital plane. Finally, the SNR for a given source and a given detector is defined as~\cite{Cutler:1994ys}
\begin{equation}
    \label{eq:SNR}
    \textrm{SNR}^2 =  4 \int_{f_{\rm m}}^{f_{\rm M}} \frac{\textrm{Re} [h^* h]}{S_n(f)} \; \textrm{d} f   \, ,
\end{equation}
where $S_n(f)$ is the detector strain sensitivity (see~\cref{sec:detectors}), while $f_{\rm m}$ and $f_{\rm M}$ are the minimal and maximal detector frequencies. Notice that in this equation, $h$ depends on all the parameters listed in~\cref{eq:signal_parameters}, but for a large sample of sources, only their average values matter in our analyses, at least at the leading order. This is why in several evaluations, \emph{e.g.}, the analytical estimates, we can average over both the spins ($\chi_1, \chi_2$) and angular ($\theta, \phi, \iota, \psi$) variables; we dub this approximated SNR as $\rm SNR_{\rm avg}$. However, as a check of robustness, in a few cases, we 
test our averaged-based results with those obtained without the average approximation, and we indicate the result of this precise evaluation simply as SNR (\emph{i.e.},~without any subscript). \\
 
As a first step, to fast probe the PBHB detectability in a broad part of their parameter space, we perform a semi-analytical analysis. For this purpose, we modify~\cref{eq:N_events_binned} by including a selection effect. Specifically, we define the expected number of resolvable sources at redshift $z$ for the population Pop as
\begin{equation}
    \hat{N}^{\rm Res, Pop}_{z} \equiv  \frac{ \Delta N^{\rm Res, Pop}_{z_{m}, z_{M}} } {\Delta z} = \frac{\mathrm{T}^{Det}_{\rm Obs}}{\Delta z} \int^{z_{M}}_{z_{m}}  \, \frac{R^{\rm Pop}(z)}{1 + z} \left[\frac{\diff V_c}{\diff z}(z)\right] \int  p^{\rm Pop}(\xi | \theta) \, \theta_{\rm SNR_{\rm avg}}(\xi)\, \diff\xi \, \diff z \, ,
    %\diff m_1 \int \diff m_2 \, p^{\rm Pop}(m_1, m_2 | \theta_{\rm Mass}) \theta_{\rm SNR}(z, m1, m2)\, ,
    \label{eq:Res_sources}
\end{equation}
where, for any given detector, the selection function $\theta_{\rm SNR_{\rm avg}}(\xi)$ is a Heaviside $\Theta$ function filtering the sources with $\rm SNR_{\rm avg}$ larger 
than $\rm SNR_{\rm Thr}=8$. The integrals in~\cref{eq:Res_sources} are carried out numerically. Notice that due to the presence of the selection function, the integration over the $\xi$ variables has to be computed explicitly.\\

We use~\cref{eq:Res_sources} to set our (analytic) criterion for the identification of the 
PBHB component via resolvable sources at Earth-based detectors. We define the PBH contribution to be visible 
if, in a given bin in $z$, it satisfies the condition
\begin{equation}
    \hat{N}^{\rm Res, PBHB}_{z } >  3   \Delta^{\rm Res, Fid}_{z } \equiv 3 \,\sqrt{ \hat{N}^{\rm Res,  Fid}_{z } }  \, .  \label{eq:DetectableSubPopGroundBased}
\end{equation}
In other words, the PBHB component can be separated from the fiducial SOBHB component if there exists a bin in $z$, in which the number of resolvable sources, $\hat{N}^{\rm Res, PBHB}_{z }$, 
exceeds the number of detectable SOBHB sources, $\hat{N}^{\rm Res, SOBHB}_{z}$, by 3$\,\sigma$. In our case, the error comes from a Poissonian distribution, and this is why we have $\Delta^{\rm Res, Fid}_{z } = \sqrt{\hat{N}^{\rm Res, Fid}_{z}}$.\\

The semi-analytic analysis is fast but disregards two effects: the populations' realization dependence and the impact of the angular and spin variables on the SNR. We quantify these effects by running a more sophisticated analysis on some PBHB population benchmarks. Specifically, we use the code in ref.~\cite{SynthesisCodePaolo} to sample over the PDF in~\cref{eq:GeneralSourceGenerator} and generate $n_{\rm r}=100$ catalogs of the SOBHB fiducial population and one catalog per PBHB benchmark population. Then, for every merger event predicted in the catalogs, we compute $\rm SNR_{\rm avg}$ and the exact SNR. We define as $\Delta \hat{\mathcal N}^{\rm Res, Pop}_{z_{\rm m}, z_{\rm M} }$ the number of events with $\rm SNR_{\rm avg}>SNR_{Thr}$ in a given redshift interval $z_{\rm m} \leq  z \leq z_{\rm M} $ for the catalogue $i$ of the population Pop. Similarly, we define as $\Delta \overline{\mathcal N}^{\rm Res, Pop}_{z_{\rm m}, z_{\rm M, i} }$ the analogous quantity obtained with the detectability condition $\rm SNR > SNR_{Thr}$. For convenience, we also introduce 
\begin{equation}
\hat{\mathcal N}^{\rm Res, Pop}_{z, i} \equiv \Delta \hat{\mathcal N}^{\rm Res, Pop}_{z_{\rm m}, z_{\rm M},i }/\Delta z \; ,\qquad\qquad  
    \overline{\mathcal N}_{z, i}^{\rm Res, Pop} \equiv \Delta \overline{\mathcal N}^{\rm Res, Pop}_{z_{\rm m}, z_{\rm M},i }/\Delta z\; .
\end{equation}

The mismatches between $\hat{\mathcal N}^{\rm Res, Pop}_{z, i}$ and  $\hat{N}_{z}^{\rm Res, Pop}$ highlight the effect of the realization dependence that our semi-analytic results neglect. However, we expect the mismatch to be statistically within the Poisson deviation from the mean, \emph{i.e.},
\begin{equation}
\left| \hat{\mathcal N}^{\rm Res, Pop}_{z, i} -  \hat\mu^{\rm \rm Res, Pop}_z  \right| < 
    3 \,\hat\sigma^{\rm Pop, Res}_z   
    \qquad \textrm{at 95\% C.L.}\;,
    \label{eq:comp_realization}
\end{equation}
where
\begin{equation}
   \hat\mu^{\rm \rm Res, Pop}_z \equiv 
        \frac{ \Delta \hat\mu ^{\rm Res, Pop}_{z_{\rm m}, z_{\rm M} } }{ \Delta z} \equiv  
             \frac{1}{n} \sum_{i=1}^{n} \frac{ \Delta \hat{\mathcal{N}}^{\rm Res, Pop}_{z_{\rm m}, z_{\rm M}, i } }{\Delta z} \, ,
   \label{eq:FidAvgResSources}
\end{equation}
\begin{equation}
\hat\sigma^{\rm Pop, Res}_z = \sqrt{\frac{1}{\Delta z}  
   \sum_{i = 1}^{n} 
        \frac{\left[ \Delta \hat{\mathcal{N}}^{\rm Res, Pop}_{z_{\rm m}, z_{\rm M}, i }  - \Delta \hat\mu ^{\rm Res, Pop}_{z_{\rm m}, z_{\rm M} } \right]^2}{n - 1}} \, .
\label{eq:SigmaResSources}
\end{equation}  
Here, the index $i$ runs over $n$, the number of realizations of each population scenario. Since we produce multiple realizations only of our fiducial population (namely $n=n_r=100$), we focus on the realizations of this population to test~\cref{eq:comp_realization}. This allows us to prove $\sigma^{\rm Res, Pop}_z\simeq \Delta^{\rm Res, Pop}_z$ and, in turn, to use $\Delta^{\rm Res, Pop}_z$ as a proxy of the realization dependences in our PBHB benchmarks.\\

The quantities $\overline{\mathcal{N}}^{\rm Res, Pop}_{z_{\rm m}, z_{\rm M}, i }$ are 
useful to investigate the impact of the approximation $\rm SNR_{\rm avg}$, in which the SNR is computed by averaging over the angles and spin. For this purpose, we calculate the quantities $\overline{\mu}^{\rm \rm Res, Pop}_z$ and  $\overline{\sigma}^{\rm Pop, Res}_z$, given as in eqs.~\eqref{eq:FidAvgResSources} and \eqref{eq:SigmaResSources} but with the hat symbol replaced by the bar one. In the parameter regions where the approximation is satisfactory,  $\hat{\mu}^{\rm Res, Pop}_z$ and $\overline{\mu}^{\rm Res, Pop}_z$ are expected to be practically equal\footnote{In principle, it is possible to replace the whole semi-analytic approach with the much more time-consuming method based on realizations and precise SNR evaluation. In this case, the detection criterion for a PBH population realization would be $\overline{ \mathcal{N}}^{\rm Res, Benchmark}_{z}   > 3 \, \overline{\sigma}^{\rm Fid, Res}_{z } $ instead of \cref{eq:DetectableSubPopGroundBased}.}. All these quantities can be computed for different values of $\rm T^{\rm Det}_{\rm Obs}$ and several detector sensitivities. For concreteness, we consider LIGO A${}^{\rm +}$ and ET, assuming $T^{\rm Det}_{\rm Obs} =1, 10$\,yr of data. Moreover, to simplify the notation, hereafter we drop all the $z$ subscripts in all these quantities and refer to the quantity, say $q$, as $q^{\rm Res, Pop}$.\\

Let us comment on the assumptions of our semi-analytic analysis. The most relevant assumption pertains to the procedure to evaluate the right-hand side of~\cref{eq:DetectableSubPopGroundBased}. In particular, while a consistent analysis should account for both the realization error and the uncertainty in the model parameters, we evaluate~\cref{eq:DetectableSubPopGroundBased} using the central values for all the fiducial population parameters without including their uncertainties. There are three main reasons behind this choice:
\begin{enumerate}
    \item The main message of this work is to stress the synergy between Earth-based and space-based GW detectors for what concerns assessing the presence of populations of high-redshift SMBHBs beyond our fiducial population. Thus, including further uncertainties in the analysis will quantitatively affect our results, but it will not change the message of the present work.
    \item Current GW detections have only probed the Universe at $z \lesssim 1$ so that the peak in the SOBHB population directly descends from imposing the population to follow the SFR at high redshift. Moreover, we have no information on the possible presence of time delays, which might shift the SOBHB peak position. All these uncertainties should also be included to perform a consistent analysis.
    \item With more measurements to come in the next few years (with improved sensitivity and possibly with more detectors joining the existing network), the determination of the model parameters will improve significantly. Currently, we have no reliable, precise estimate on the errors in the measured values of the fiducial population parameters at the end of the next LVK runs\footnote{Data for the O4 run, which has both longer acquisition time and better sensitivity compared to O3, will be released in the next couple of years~\cite{Cahillane:2022pqm, Kiendrebeogo:2023hzf}, leading to significant improvement in the determination of all the population parameters. The parameter determination will improve even further with O5.}.
\end{enumerate}
For these reasons, we restrict ourselves to the case where the main uncertainty on $\hat{N}^{\rm Res, Pop}$ comes from the Poissonian error. However, we show that this treatment is reasonably good by performing a more computationally demanding analysis on some benchmarks. We postpone the exhaustive treatment of the realization dependencies and population uncertainties to the time when more LVK data will be available. 

\subsection{SGWB analysis}
\label{sec:SGWBFishMat}

The SGWB from the fiducial SOBHB population, and its variation due to the PBHB contribution, is evaluated using the analytical approach summarized in~\Cref{app:analytical_SGWB} and detailed in ref.~\cite{Phinney:2001di}. It turns out that, within the LISA frequency band, the SGWB signal sourced by each population can be parametrized as a simple power-law:
\begin{equation}
h^2 \Omega^{\rm Pop}_{\rm GW}(f) = 10^{\alpha_{\rm Pop}} \left( \frac{f}{f_{*} }\right)^\beta \;,
\label{eq:power_law}
\end{equation} 
where $\beta=2/3$ is the tilt of the GW power spectrum and $\alpha_{\rm Pop}$ is the ($\log_{10}$ of) the amplitude at a reference (irrelevant) pivot frequency $f_{*}$. This result assumes each frequency bin to be highly populated by the GW signals due to binaries in circular orbits, with negligible environmental effects, and emitting GWs only~\cite{Cusin:2019jhg, PhysRevD.103.043002, Bavera:2021wmw, next}. Dropping any of these assumptions might induce modifications from the power-law behavior\footnote{Also the eccentricity of the orbit, astrophysical uncertainties, or individual source subtraction might affect the shape of the SGWB generated by a population of compact objects~\cite{Rajagopal:1994zj, Jaffe:2002rt, Wyithe_2003, Cornish:2017vip, DOrazio:2018jnv, Zhao:2020iew, Karnesis:2021tsh, Lehoucq:2023zlt, Babak:2023lro}. \label{footnote:SGWB_shape}}. 
Since we are interested in evaluating the SGWB in the LISA frequency band  $f \in \{3 \times 10^{-5} \,, 0.5 \}$\,Hz, it is convenient to set $f_* = 0.01$\,Hz. \\

Our strategy to probe the presence of the PBHB population via the SGWB measurement consists of reconstructing the overall signal with the above power-law template, and then checking whether the posteriors of the template parameters are compatible with those predicted by the fiducial SOBHB population up to some confidence level. As the aforementioned assumptions imply $\beta=2/3$ for both SOBHB and PBH populations, only deviations from the SOBHB amplitude parameter are relevant to our goal. We calculate the reference SOBHB population amplitude
by using \cref{eq:SGWB_hcFinal} and \cref{eq:SGWB_OmegaUnits} where, for practical purposes, the integral over $z$ is cut off above a maximal redshift. Specifically, for the SOBHB population, we integrate up to $z \approx 10$ and obtain $\alpha_{\rm SOBHB} \simeq -12.02$ at $f_* = 0.01$Hz. With such a maximal-redshift choice, the estimate of the amplitude is accurate to within $1 \%$ error~\cite{Babak:2023lro}. We follow the same procedure to compute $\alpha_{\rm PBH} $ predicted by a PBH population with a given set of hyperparameter values. However, the (unresolved) PBHB signals do not die off as fast as the SOBHB ones. For this contribution, we set the cut-off at $z=100$ corresponding to a $\sim10\%$ accuracy in the SGWB evaluation\footnote{To achieve the 1\% accuracy level in the computation of $\alpha_{\rm PBH} $, one would have to integrate much higher values of $z$. Given the theoretical uncertainties on the distribution of PBHB at such high redshifts, we choose a cut-off that reasonably compromises between numerical and theoretical uncertainties~\cite{LISACosmologyWorkingGroup:2023njw}.}. \\

To forecast the LISA posteriors on the power-law template parameters, we perform an analysis based on the Fisher Information Matrix (FIM) formalism. Given some data $\tilde{d}(f)$, containing signal $\tilde{s}(f)$ and noise $\tilde{n}(f)$, which we assume to be Gaussian, with zero means, and characterized only by their variances\footnote{In reality, the resolution $\Delta f$ is finite and given by $1 / T$. As long as this frequency is much smaller than $f_{\rm min}$, we can effectively replace the discrete sums with integrals.}, the (log-)likelihood can be written as
\begin{equation}
- \log \mathcal{L} (\tilde{d} \vert \vec{\theta} ) \propto T \int_{f_{\mathrm{min}}}^{f_{\mathrm{max}}}  \left\{ \ln \left[ D(f, \vec{\theta}) \right] + \frac{ \tilde{d}(f) \tilde{d}^* (f) }{  D(f, \vec{\theta}) }   \right\}\; \textrm{d} f \; , 
\end{equation}
where $f_{\mathrm{min}}$ and $f_{\mathrm{max}}$ are the minimal and maximal frequencies measured by the detector, $T$ is the total observation time, and $D(f, \vec{\theta})$ is the model for the variance of the data, depending on some (signal and noise) parameters $\vec{\theta}$. The best-fit parameters $\vec{\theta}_0$ are defined to maximize $\log \mathcal{L}$:
\begin{equation}
   \left. \frac{\partial \log \mathcal{L} }{ \partial \theta^\alpha  } \right|_{\vec{\theta} =  \vec{\theta}_0 } \propto  T \int_{f_{\mathrm{min}}}^{f_{\mathrm{max}}}  \frac{ \partial \ln D(f, \vec{\theta})}{\partial \theta^\alpha} \left[ 1 - \frac{ \tilde{d}(f) \tilde{d}^{*}(f) }{D(f, \vec{\theta})} \right]   = 0  \; ,
\end{equation}
which is clearly solved by $D(f, \vec{\theta}_0) = \tilde{d}(f) \tilde{d}^{*}(f)$. Then, the FIM $F_{\alpha \beta}$ is given by
\begin{equation}
\label{eq:FIM_definition}
F_{\alpha \beta} \equiv - \left. \frac{\partial^2 \log \mathcal{L} }{ \partial \theta^\alpha \partial \theta^\beta } \right|_{\vec{\theta} =  \vec{\theta}_0} =  T \int_{f_{\mathrm{min}}}^{f_{\mathrm{max}}} \frac{\partial \log D(f, \vec{\theta})}{\partial \theta^\alpha} \frac{\partial \log D(f, \vec{\theta})}{\partial \theta^\beta} \, \textrm{d}f \; .
\end{equation}
By definition, the FIM $F_{\alpha \beta}$ is the inverse of the covariance matrix $C_{\alpha \beta}$. As a consequence, estimates of the errors on the model parameters $\vec{\theta}$  are obtained by computing $\sqrt{ \textrm{diag}(C_{\alpha \beta}) }  = \sqrt{ \textrm{diag}(F_{\alpha \beta}^{-1}) }$. Notice that LISA will measure three data streams. Under some simplifying assumptions, these data streams in the AET TDI basis are independent (see~\cref{sec:detectors}), and therefore the total Fisher matrix is given by
\begin{equation}
    F^{\rm Tot}_{\alpha \beta} \equiv F^{\rm AA}_{\alpha \beta}  +F^{\rm EE}_{\alpha \beta}  +F^{\rm TT}_{\alpha \beta} = 2 F^{\rm AA}_{\alpha \beta} + F^{\rm TT}_{\alpha \beta} \; .
\end{equation}
For what concerns the observation time $T$, we assume $100 \%$ efficiency and impose $T^{ \rm LISA}_{\rm Obs} = 4$\,yrs. For reference, we also show how results improve if the mission lifetime is extended to $10$\,yrs. Let us assume that the data are expressed in $\Omega$ units, and we have factored the detector response out (for details, see~\cref{sec:detectors}). Then, $D(f, \vec{\theta})$ can be expanded as
\begin{equation}
D(f, \vec{\theta}) = h^2 \Omega_{\rm GW}(f, \vec{\theta}_s) +  h^2 \Omega_{n}(f, \vec{\theta}_n) \; ,
\label{eq:FshMtrx_Df} 
\end{equation}  
where $\Omega_{\rm GW}(f, \vec{\theta}_s)$ is the template of the signal as a function of the frequency and the parameters $\vec{\theta}_s=\{\beta, \alpha_{\rm Tot}\}$, and $\Omega_{n}(f, \vec{\theta}_n)$ is the noise model as a function of the frequency and the noise parameters $\vec{\theta}_n=\{A,P \}$. For what concerns the noise, we use the analytical two-parameter model commonly used in the literature (for details, see~\cref{sec:detectors}). \\

In our case,  the signal is a sum of two contributions, one for the fiducial population and one for the PBHB population, each described by the template in~\cref{eq:power_law}. Given the complete degeneracy between these two components, we cannot measure them independently, but only the overall amplitude, which in the pivot frequency is given by $\alpha_{\rm Tot} = \log_{10} (10^{\alpha_{\rm Fid}} + 10^{\alpha_{\rm PBH}}) $. In particular, to assess the significance of the PBHB contribution, we check whether, for each PBHB population, the value of $\alpha_{\rm Fid}$, with its error band, is not compatible at some $\sigma$-level (from 1 to 3) with $\alpha_{\rm Tot}$, \emph{i.e.},
\begin{equation}
    \alpha_{\rm Tot} - n \,\sigma_{\alpha, \rm Tot} > \alpha_{\rm Fid} + n\, \sigma_{\alpha, \rm Fid} \; ,  
    \label{eq:SGWB_condition}
\end{equation}
where $\sigma_{\alpha, \rm Tot}$ and $\sigma_{\alpha, \rm Fid}$ are the FIM errors on $\alpha_{\rm Tot}$ and $\alpha_{\rm Fid}$, respectively, and $n \in \{ 1,2,3\}$. For reference, we report that for $\alpha_{\rm Fid} = \alpha_{\rm SOBHB} \simeq -12.02$ at $f_* = 0.01$Hz, we have $\sigma_{\rm \alpha, Fid}^{4 \rm yr} \simeq 8.44 \times 10^{-3}$ and $\sigma_{\alpha , \rm Fid}^{10 \rm yr} \simeq 5.34 \times 10^{-3}$ at 68\% confidence level after marginalizing over the error on $\beta$. Such findings are compatible with those of ref.~\cite{Babak:2023lro}\\

As in the previous section, we conclude by discussing the limitations of our analysis. Analogously to Earth-based detectors and following similar lines of reasoning, we do not include the uncertainties in the population parameters in our analyses\footnote{A detailed study on how these uncertainties would affect the SGWB signal at LISA, can be found in ref.~\cite{Babak:2023lro}.}. A further crucial approximation is that the FIM formalism gives accurate estimates for the uncertainties in determining the model parameters. This approximation holds in the limit where the log-likelihood for the model parameters is sufficiently Gaussian around $\vec{\theta}_0$, which should be quite accurate for the specific injections considered in this work~\cite{Babak:2023lro}.

\section{Results and discussion}
\label{sec:Results}

\begin{table}[htb!]
\centering
\begin{tabular}{|c|c|}
\hline
Parameter & Range \\ \hline
$R_0$ fraction & $\varepsilon \in [10^{-3}, 1]$ \\ \hline
Mass PDF central parameter & $ \mu_{\rm LN} \in [0, 100]$ \\ \hline
Mass PDF standard deviation & $ \sigma_{\rm LN} = [0.1, 0.5, 1, 2.5]$ \\ \hline
Integrated mass range & $ m \in [0, 150]$ \\ \hline
Earth-based integrated redshift range & $ z \in [0, 10]$ \\ \hline
SGWB integrated redshift range & $ z \in [0, 10^{2}]$ \\ \hline
\end{tabular}
\caption{The range of parameter values used in the semi-analytic analysis.}
\label{tab:AnalysisPrior}
\end{table}

We now present the results obtained by applying the proposed methodology to the PBHB population scenario introduced in \cref{sec:PBHSubPopMod}. We describe the semi-analytic results achieved by considering the LISA and LIGO $\rm A^+$ detectors in \cref{sec:PBHMapsLIGO}, while those obtained with LISA and ET in \cref{sec:PBHMapsET}.
Given these results, we select a set of benchmark points in the PBHB population parameter space. On these benchmarks, we assess the robustness of the semi-analytic analysis by incorporating spin and sky location parameters, as well as considering realization dependence effects. The results of this assessment are presented in \cref{sec:PBHBenchAnal}. The overall analysis is conducted on the PBHB parameter space summarized in \Cref{tab:AnalysisPrior}, assuming $1$ and $10$ years of continuous observations for the Earth-based detectors, and $4$ and $10$ years of continuous measurements for LISA.

\subsection{Detectability of PBHB populations using LISA and LIGO A${}^+$}
\label{sec:PBHMapsLIGO}

\begin{figure}[htb!]
  \centering
   \begin{subfigure}[b]{\textwidth}
   \includegraphics[width=\columnwidth]{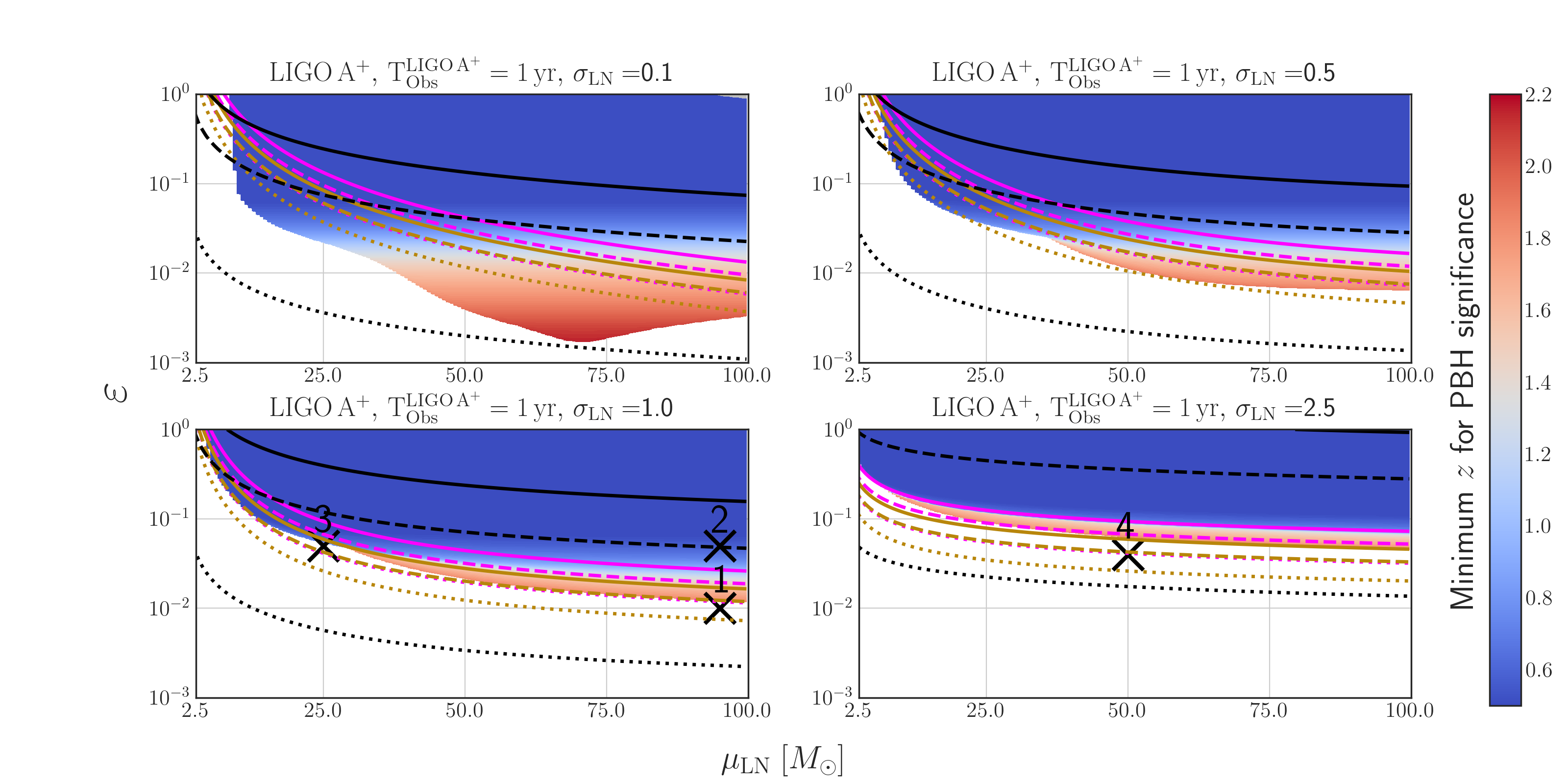}
   \end{subfigure}
   \begin{subfigure}[b]{\textwidth}\includegraphics[width=\columnwidth]{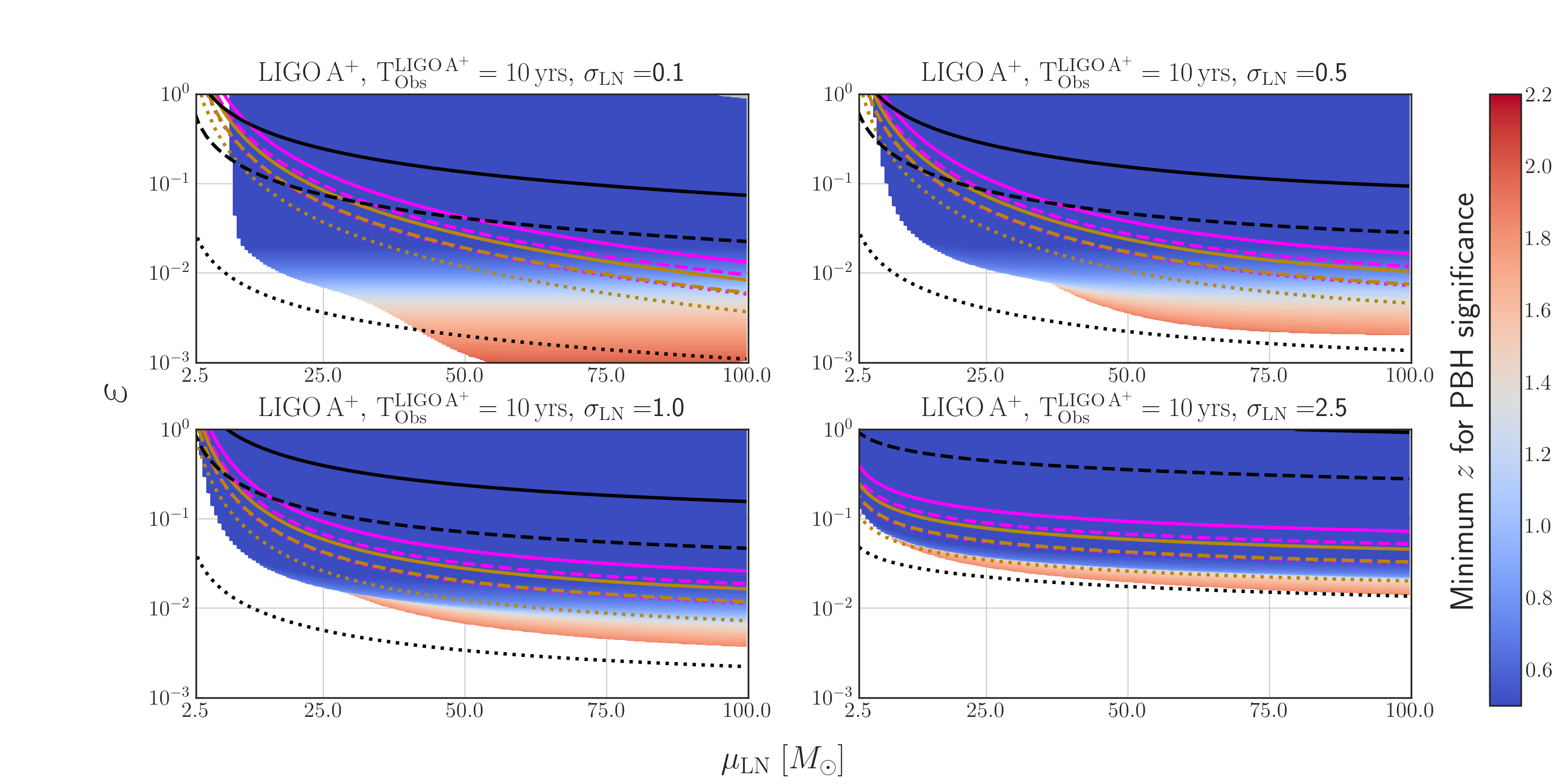}
   \end{subfigure}
   \begin{subfigure}[b]{\textwidth}
   \centering
   \includegraphics[width=0.75\columnwidth]{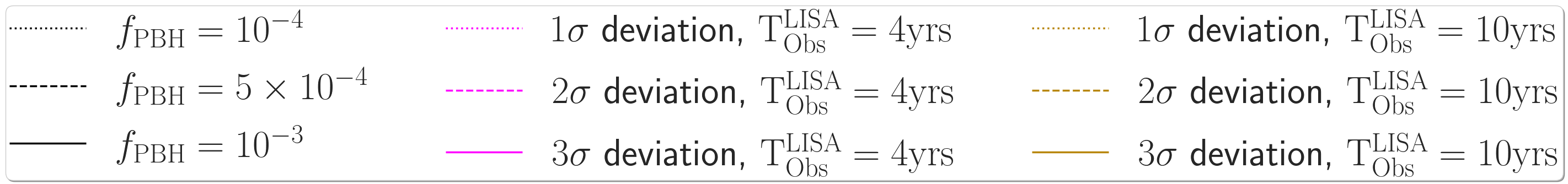}
   \end{subfigure}
\caption{LISA and LIGO A${}^{+}$ prospect for the detection of LN PBHB population in the parameter space plane $\{\varepsilon, \mu_{\rm LN}\}$. Values of the peak width, $\sigma_{\rm LN}$, and the number of years of LIGO A${}^{+}$ data, $\rm T^{\rm LIGO \, A^+}_{\rm Obs}$, are specified in the title of each panel.  In the regions above the magenta 
(dash-dotted) [solid] line, the amplitudes of the overall SGWB and the fiducial model measured by LISA with 4  years are incompatible at 1 (2) [3] $\sigma$ level. Light brown lines represent the same but for 10 years of LISA data. The color map indicates the minimal value of $z$ at which the condition in~\cref{eq:DetectableSubPopGroundBased} is satisfied. Crosses indicate the benchmark points investigated in our dedicated analyses. The black lines correspond to different values of $f_{\rm PBH}$. \label{fig:PBHMapsLNLIGO}}
\end{figure}

\Cref{fig:PBHMapsLNLIGO} summarizes the outcome of the semi-analytic analysis showcasing the synergy between LIGO A${}^+$ and LISA in the slice $\{\mu_{\rm LN},\varepsilon\}$ of the PBHB parameter space. The third, free parameter of the PBHB model, $\sigma_{\rm LN}$,  is set as specified in the title of each panel, which also includes the value of $\rm T^{\rm LIGO \, A^+}_{\rm Obs}$ adopted in the LIGO A${}^+$ resolvable-event measurement. 
%in the case when assuming either $T_{\rm Obs} = 1, \, 10 \, \rm yrs$ of data (top and bottom subfigure, respectively). Each panel of these subfigures corresponds to a different value for $\sigma_{\rm LN}$ introduced in~\cref{eq:PBH_LNMassPDF}. On the other hand, the $x$ axes of these plots span different values for $\mu_{\rm LN}$, and the $y$ axes correspond to different values of $\varepsilon$ defined in~\cref{eq:PBH_Rtz}. 
For reference, each panel includes the curves corresponding to $f_{\rm PBH} = 10^{-4}, \, 5 \times 10^{-4}, \, 10^{-3}$ (see~\Cref{app:ConvMaps}) marked as indicated in the legend\footnote{As these lines prove, we are far away from the region with $f_{\rm BBH}\sim 1$, for which several experimental constraints exist \cite{EROS-2:2006ryy, Ricotti_2008, Wyrzykowski_2011}. Further bounds on the abundance of stellar-mass PBHs can be found in refs.~\cite{Garcia-Bellido:2017xvr, Khalouei:2020bim, LISACosmologyWorkingGroup:2023njw}.
Since in the present paper, we are more interested in the methodology than the illustrative LN PBHB population application, we do not recast any PBH (model dependent) bounds here. In any case,  these bounds should not rule out the bulk of the considered parameter with $f_{\rm BBH} \lesssim 10^{-3}$. }.
For each parameter point, we compute the integral in~\cref{eq:Res_sources} and look for the smallest value of $z$, say $\bar{z}$, such that the condition in~\cref{eq:DetectableSubPopGroundBased} is satisfied. The value of $\bar{z}$ sets the color in all these plots. White areas correspond to the parameter regions where the condition in~\cref{eq:DetectableSubPopGroundBased} is not satisfied at any $z$. In the colorful region, therefore, LIGO A${}^+$ would recognize the presence of the PBHB component in the SMBHB population (within the assumptions of the semi-analytic analysis). Moreover, LISA would reach the same conclusion in the PBH parameter regions above the magenta and light brown lines, which are computed by evaluating the condition in~\cref{eq:SGWB_condition} (see the legend for the LISA observation time, $\rm T^{\rm LISA}_{\rm Obs}$, and the sigma level, $n$, corresponding to each line).  In the tiny (almost not visible) gray areas appearing in the top right corner of some panels, the overall SGWB signal violates the current LVK upper bound on the SGWB amplitude~\cite{KAGRA:2021kbb}. Finally, the crosses are the benchmark points that we select for the tests in~\cref{sec:PBHBenchAnal}. 
\\

First of all, by scrutinizing~\Cref{fig:PBHMapsLNLIGO}, we understand that LIGO $\rm A^+$ will observe events from the LN PBHB populations only up $z \lesssim 2$. This is both a consequence of the detector's sensitivity, which only allows for detecting events up to $z \lesssim 3$ (see~\Cref{sec:detectors}), and of the averaging over the spin and angular variables, which slightly lowers the maximal redshift reach compared to, \emph{e.g.}, an optimally located source. Thus, LIGO $\rm A^+$ will not be able to resolve events in the range where the PBHB population naturally dominates over the SOBHB population, \emph{i.e.}, at redshift higher than the SFR peak. As a consequence, either the fraction 
%$\varepsilon$ 
of the PBHB population is relatively high at low redshift, or LIGO $\rm A^+$ will not be able to detect its presence. The additional information provided by the SGWB amplitude measured by LISA  might provide an invaluable tool to break the degeneracy among different population models. In particular, the SGWB measurement proves to be quite effective for probing models predicting very narrow peaks at low masses or very broad peaks with small values of $\varepsilon$. On the other hand,  for sufficiently large values of $\mu_{\rm LN}$, populations with narrow mass distributions are more easily detectable with Earth-based detectors. The motivation is that the SNR decreases for increasing mass ratio ($q = m_1/m_2$). For narrow mass distributions, the two PBHs of each binary are more likely to have similar masses, which, on average, increases the typical event SNR. 
Moreover, PBHB populations with extremely narrow mass distributions, located at either too small or too large masses, will not be detectable since they generate signals that are either too feeble (the GW amplitude grows with the mass of the binary) or outside the detector's frequency band (higher masses generally coalesce at lower frequencies). Examples of these effects are visible in, \emph{e.g.}, the left and right top panels of~\Cref{fig:PBHMapsLNLIGO}. In particular, in the top left panel, the minimum in the border of the colorful area  at $\mu_{\rm LN} \simeq 70 M_{\odot}$ originates from the interplay between these two effects. Notice also that the shape of the border of the colorful region before such minimum tracks the behavior of the fiducial mass function, which, as discussed in~\Cref{app:FidMod}, is assumed to be a \emph{power-law + peak model}~\cite{KAGRA:2021duu}. Such a shape is expected in the limit  $\sigma_{\rm LN}\to 0$, and gets smoothed out by increasing $\sigma_{\rm LN}$, as~\Cref{fig:PBHMapsLNLIGO} shows. \\

Comparing the panels with $\rm T^{\rm LIGO \, A^+}_{\rm Obs} = 1$ year with those with $\rm T^{\rm LIGO \, A^+}_{\rm Obs} = 10$ years is also interesting. Firstly, we see that, for Earth-based detectors and LISA, increasing the effective observation time generally enhances the detection prospects but not significantly affect the qualitative behavior of the results. Indeed, for Earth-based detectors, increasing the observation time does not affect the single events detection, but rather, it only increases their number, improving the overall statistics. Similarly, for LISA, increasing the observation time does not impact the overall SGWB amplitude\footnote{As already mentioned in~\cref{footnote:SGWB_shape}, this surely holds within our assumptions but might not be generally true, as with longer observation time and/or archival searches, the individual source subtraction will improve, leading to changes both in the amplitude and shape of the SGWB.}, but rather, the accuracy of its measurement.\\

Overall, \Cref{fig:PBHMapsLNLIGO} leads to conclude that, for  PBHB mass functions peaked at masses maximizing the SNR, $\mu_{\rm LN} \simeq 70 M_{\odot}$, PBH populations with even tiny $\varepsilon$ $(\varepsilon \sim 10^{-3})$ are within the reach of LIGO $\rm A^+$. While this quickly degrades as PBHB mass functions broaden, this behavior is not as marked for the prospect of SGWB detection with LISA. As a consequence, parts of the parameter space that can be hardly probed with Earth-based detectors can still be accessed with LISA. \\

\subsection{Detectability of PBHB populations using LISA and ET}
\label{sec:PBHMapsET}
\begin{figure}[htb!]
  \centering
   \begin{subfigure}[b]{\textwidth}
   \includegraphics[width=\columnwidth]{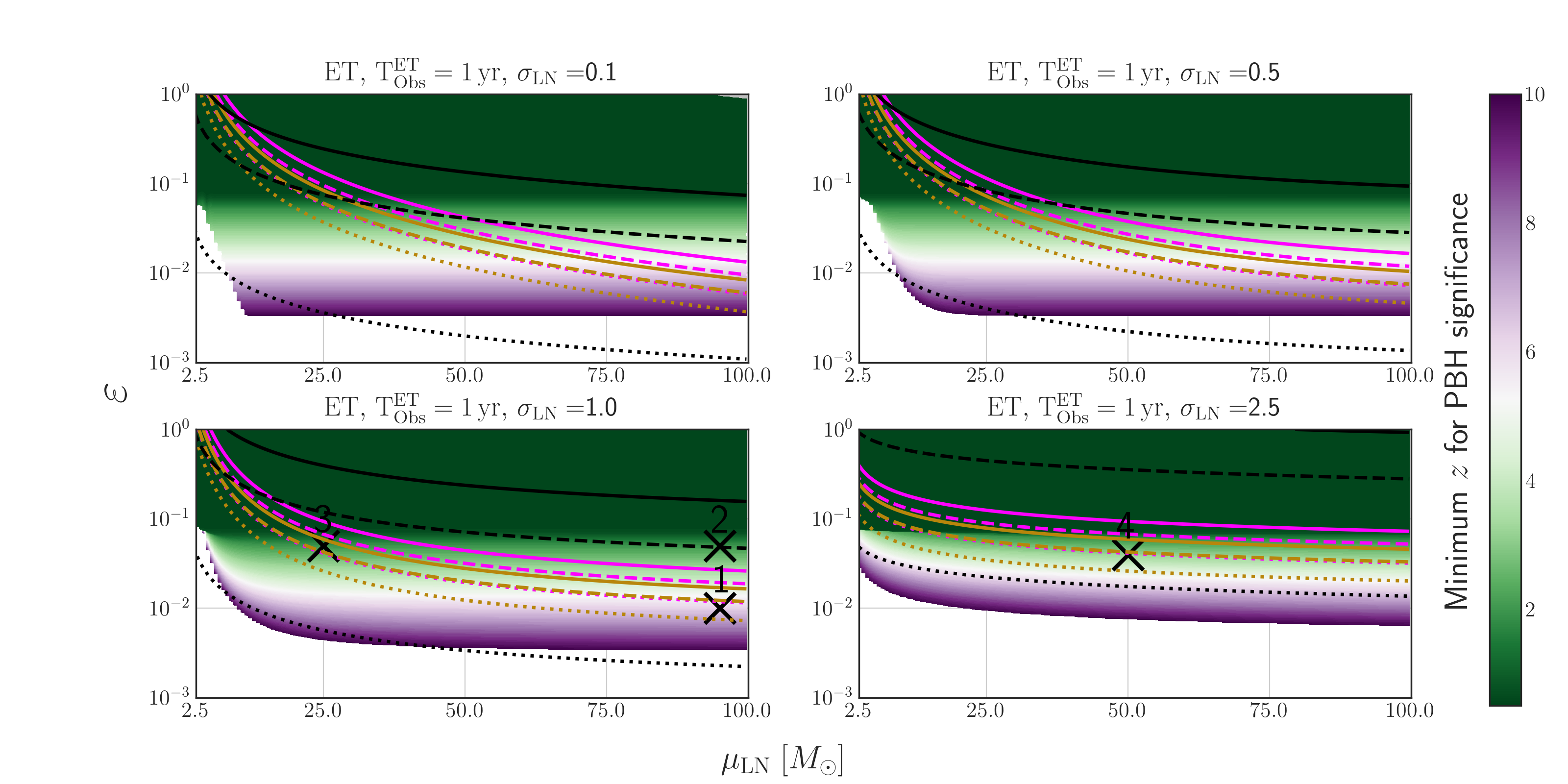}
   \end{subfigure}
   \begin{subfigure}[b]{\textwidth}
   \includegraphics[width=\columnwidth]{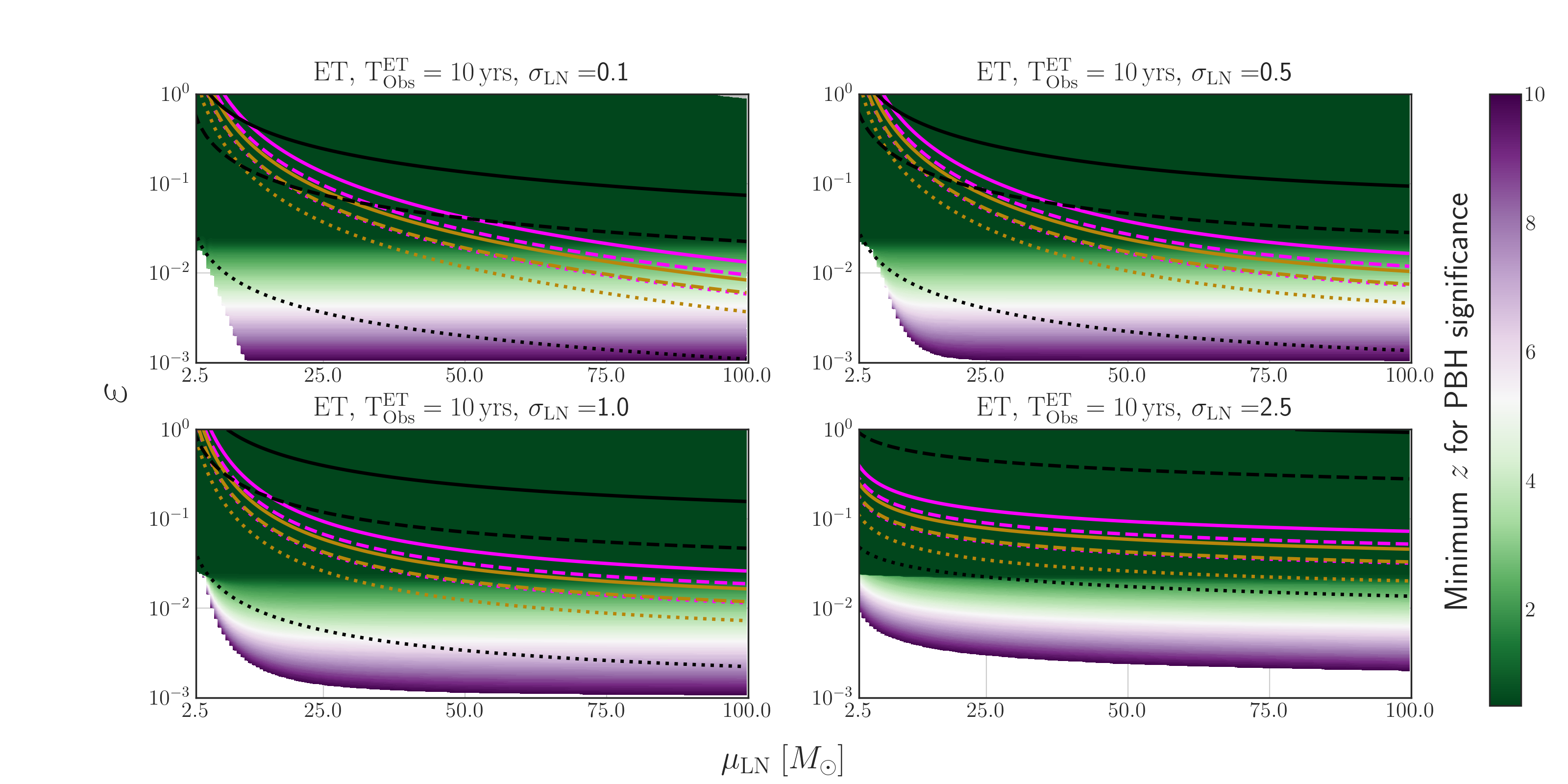}
   \end{subfigure}
   \begin{subfigure}[b]{\textwidth}
   \centering
   \includegraphics[width=0.75\columnwidth]{Figures/MapsLegend.png}
   \end{subfigure}
\caption{As in \Cref{fig:PBHMapsLNLIGO} but with ET instead of LIGO $\rm A^+$. \label{fig:PBHMapsLNET}}
\end{figure}

We proceed by discussing how the PBHB detection prospects improve when ET replaces LIGO $\rm A^+$. 
\Cref{fig:PBHMapsLNET} shows the results corresponding to this scenario. 
%
%As done for , we assume either  $T_{\rm Obs} = 1 \, \rm yr$ or $ T_{\rm Obs} = 10 \, \rm yrs$. The results for the LN mass distributions are presented in the top and bottom subfigure in~\Cref{fig:PBHMapsLNET}, respectively. 
It also displays the benchmark points identified in the previous section to highlight their detectability with ET. \\ 

As discussed in~\Cref{sec:detectors}, ET can resolve events at $z \gtrsim 10$, \emph{i.e.}, well beyond the SFR peak where the SOBHB population quickly drops. As a consequence, ET has way better prospects of identifying PBHB population departures from the SOBHB fiducial model. This fact is manifest in~\Cref{fig:PBHMapsLNET}. Moreover, by comparing \Cref{fig:PBHMapsLNLIGO} and~\Cref{fig:PBHMapsLNET}, we see that the detection prospect of ET has less dependency on $\mu_{\rm LN}$ than the one of LIGO $\rm A^+$. This effect originates from the improved sensitivity, which, for the mass and redshift ranges considered in the present work, leads to less pronounced selection effects in ET, compared to LIGO $\rm A^+$. Indeed, all panels of~\Cref{fig:PBHMapsLNET} indicate that selection effects due to the binary mass only affect the very low end of the mass range. We can thus conclude that, as long as the PBHB population will produce a sufficiently large number of events (\emph{i.e.}, larger than the Poissonian $3 \sigma$ expected for the fiducial population) at high redshift, ET will measure a significant excess in the number of events.\\

Despite the great increase in the detection ability of ET compared to LIGO A${}^{+}$, the SGWB measured by LISA still brings additional information. The main motivation for this claim is that events with very large masses (outside the range of our plots) would merge at too low frequencies to be detected with ET, but would still contribute to the SGWB amplitude. However, a similar argument could also hold for different redshift distributions predicting a few events at low redshift and many more events at very high redshift. Hence, we stress, once again, the synergy between individual events and SGWB for constraining population models.\\

While, beyond the improvements just underlined, most of the comments explained in~\cref{sec:PBHMapsLIGO} for LIGO $\rm A^+$ remain valid for ET, we remark that following the methodology introduced in~\cref{sec:IndSourc}, there are some regions where the PBHB population turns out to be detectable with smaller values of $\varepsilon$ in LIGO $\rm A^+$ than in ET (cf.~\Cref{fig:PBHMapsLNLIGO} and~\Cref{fig:PBHMapsLNET}).  This might seem counterintuitive, given that ET has better sensitivity. Indeed, this is an artefact of our choice for the detectability criterion, defined in~\cref{eq:DetectableSubPopGroundBased}, and of the selection function for the different GW detectors. With our approach, if the fiducial population produces fewer resolvable events at a given redshift, fewer events are required from the PBHB population to satisfy~\cref{eq:DetectableSubPopGroundBased}. In particular, since LIGO $\rm A^+$ selects very few events from the fiducial population, it might be easier for a PBHB population with suitable properties (\emph{i.e.}, with a narrow mass function centered at the right value to optimize the SNR at LIGO $\rm A^+$) to satisfy~\cref{eq:DetectableSubPopGroundBased}. However, a proper population analysis keeping track of both the redshift and mass distribution (see e.g.~ref.~\cite{Franciolini:2021tla}) would reveal this feature. 

\subsection{Analysis of the PBHB population benchmark points}
\label{sec:PBHBenchAnal}

\begin{table}
\centering
\begin{tabular}{|c|c|c|c|c|}
\hline
Point N. & 1 & 2 & 3 & 4\\ \hline
$\mu_{\rm LN} \, [M_\odot]$& $95.0$ & $95.0$ & $25.0$ & $50.0$\\ \hline
$\sigma_{\rm LN}$ & $1.0$ & $1.0$ & $1.0$ & $2.5$\\ \hline
$\varepsilon$ & $0.01$ & $0.05$ & $0.05$ & $0.04$ \\ \hline
LIGO $\rm A^+$ (1 yr)& N.D. & $z \sim 1$ & N.D. & N.D. \\ \hline
ET (1 yr)& $z \sim 6$ & $z \sim 2$ & $z \sim 2$ & $z \sim 3$\\ \hline
LISA (4 yrs)& $\sim 1\sigma$ & $>> 3\sigma$ & $\lesssim 2\sigma$ & $\gtrsim 1\sigma$ \\ \hline
\end{tabular}
\caption{Description of the benchmark points of the PBHB populations, the redshift at which they become notable at LIGO $\rm A^+$ and ET, and the significance of their SGWB signal at LISA. The acronym N.D. stands for non-detectable.}
\label{tab:BenchPoints}
\end{table}

In this section, we perform a more accurate analysis of the resolvable sources with LIGO $\rm A^+$ and ET focussing on the benchmark points shown in~\Cref{fig:PBHMapsLNLIGO} and~\Cref{fig:PBHMapsLNET}. The goal is to quantify the inaccuracies caused by the SNR$_{\rm avg}$ approximation and the omission of the realization dependencies. The rationale is detailed in~\cref{sec:IndSourc}.\\
%Part of this analysis includes all the waveform parameters which we have averaged over in the previous sections. 

Let us start by commenting on the choice of our benchmark points. While details are summarized in~\Cref{tab:BenchPoints}, qualitatively these points exhibit the following features:
\begin{itemize}
\item Point 2 leads to signatures in LISA, LIGO $\rm A^+$, and ET. Points 1 has the same $\varepsilon$ of Point 2 but smaller $\mu_{\rm LN}$, while Point 3 has the same $\mu_{\rm LN}$ but smaller $\varepsilon$. Both Point 1 and Point 3 are marginally detectable with LISA and LIGO $\rm A^+$, but well testable with ET.

\item Point 4 exhibits a so large $\sigma_{\rm LN}$ that its mass function is almost flat in the range of interest. It is visible at ET, barely visible at LISA, but not visible at LIGO $\rm A^+$.
\end{itemize}
\begin{figure}[htb]
  \centering
   \includegraphics[width=0.46\columnwidth]{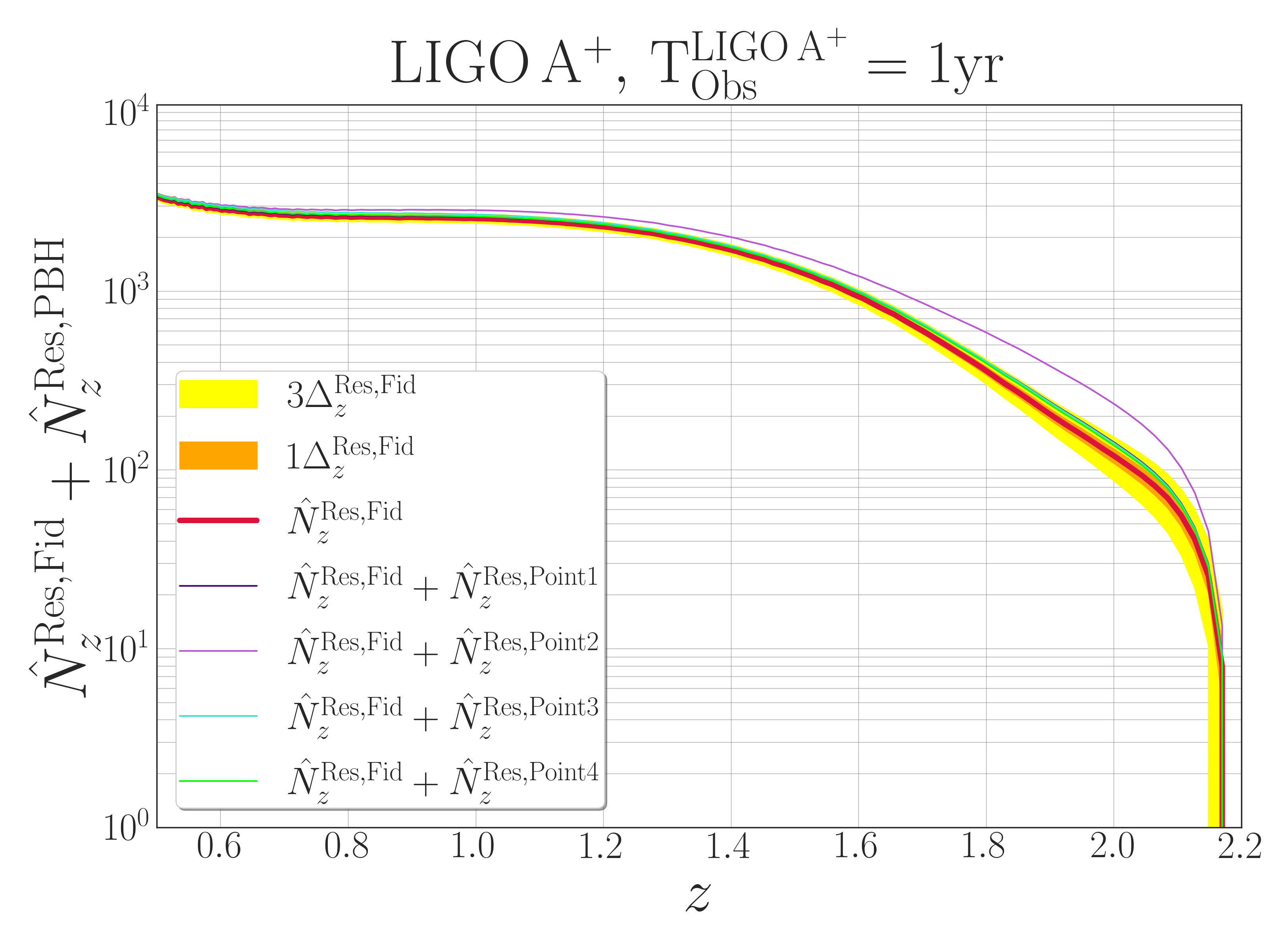}
   \includegraphics[width=0.46\columnwidth]{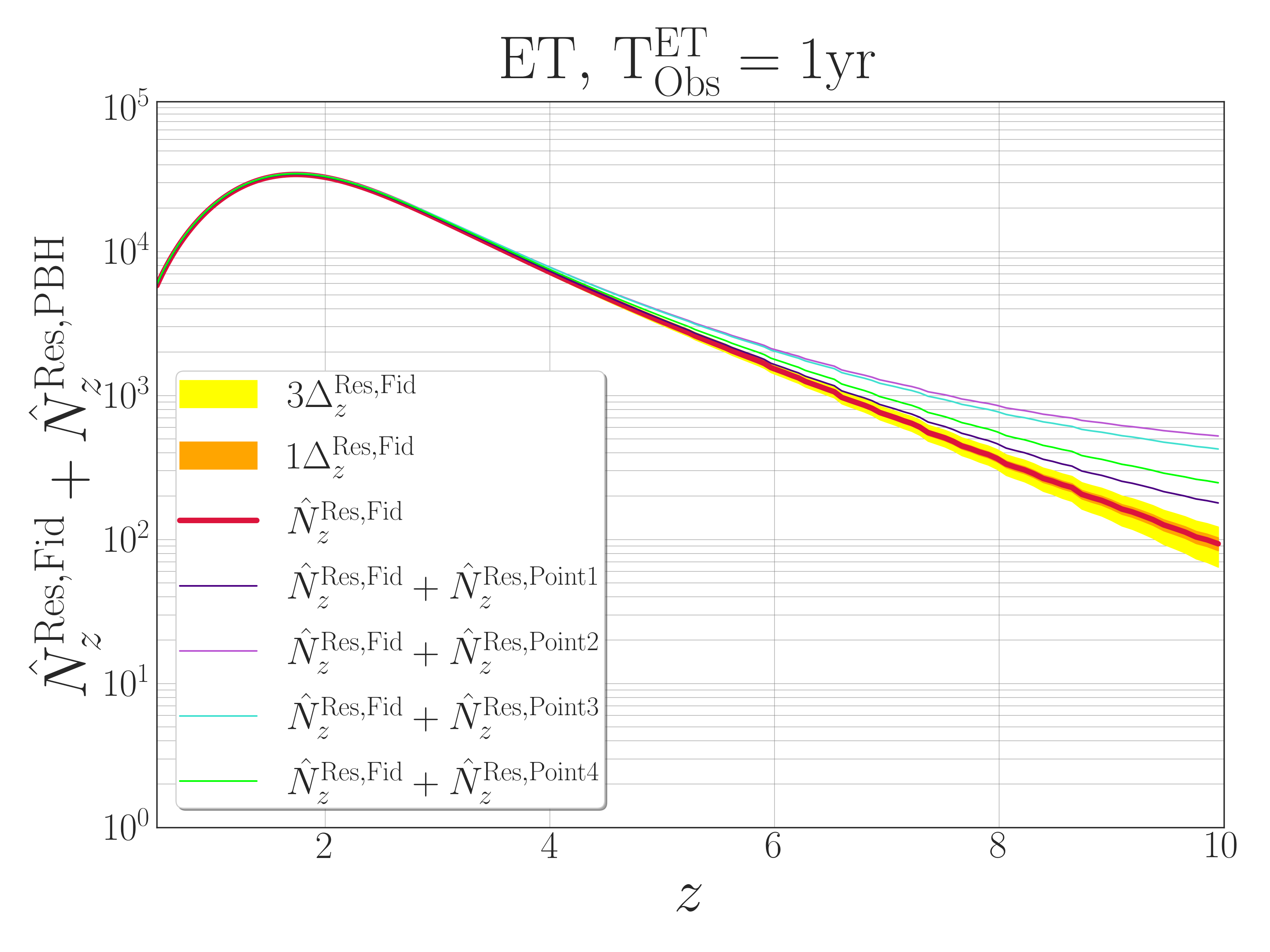}\\   
   \includegraphics[width=0.46\columnwidth]{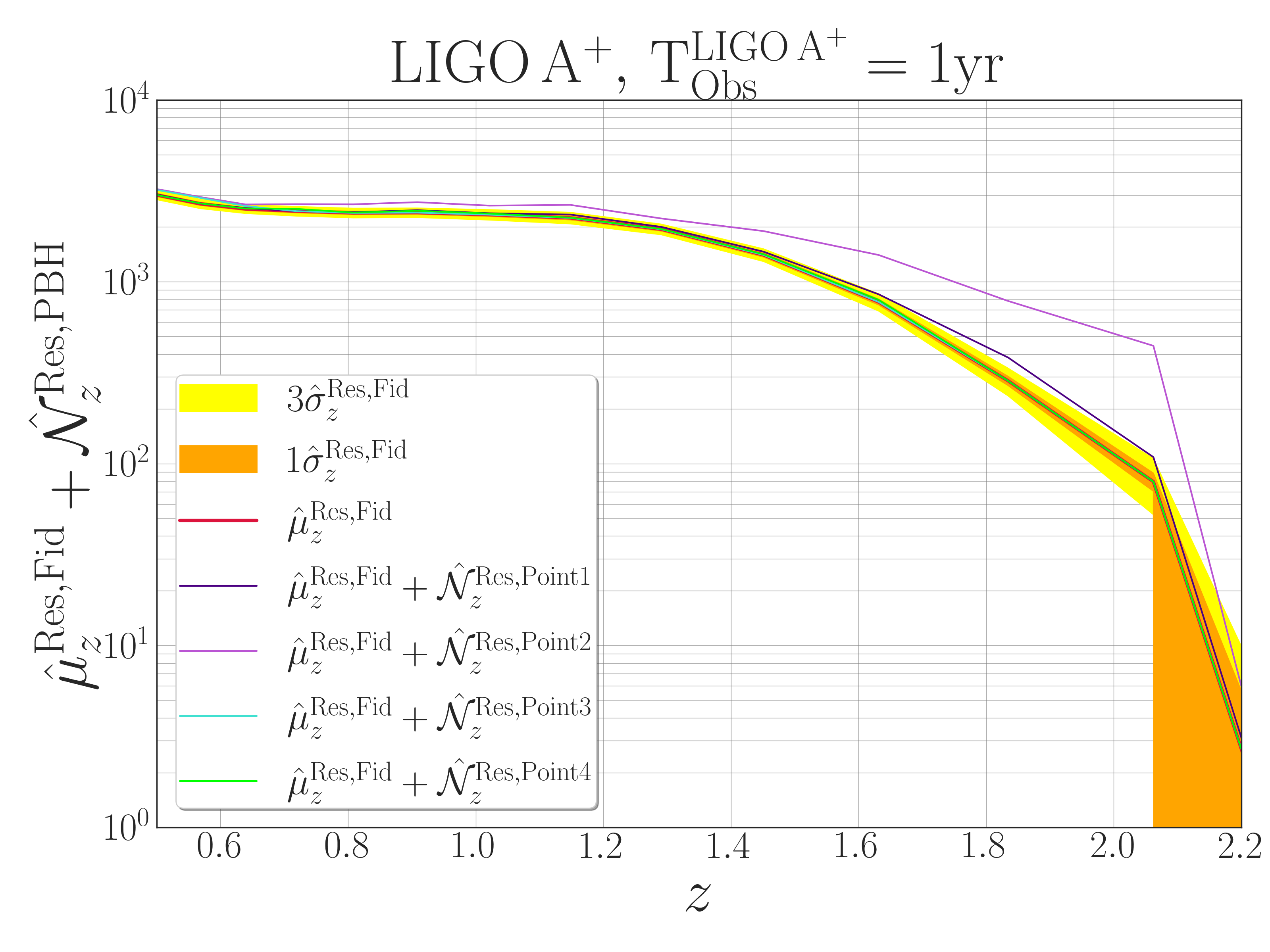}
   \includegraphics[width=0.46\columnwidth]{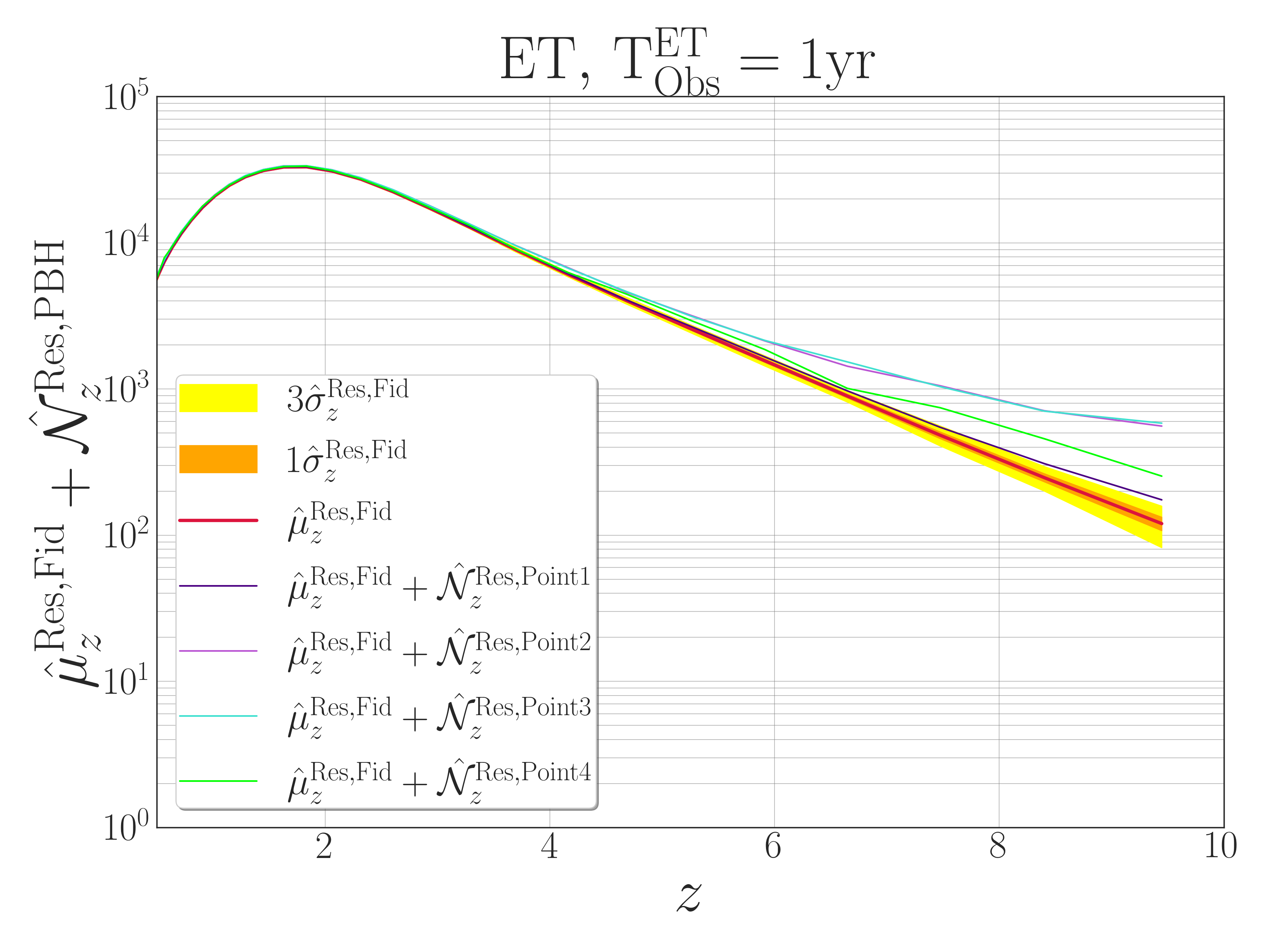}\\ 
   \includegraphics[width=0.46\columnwidth]{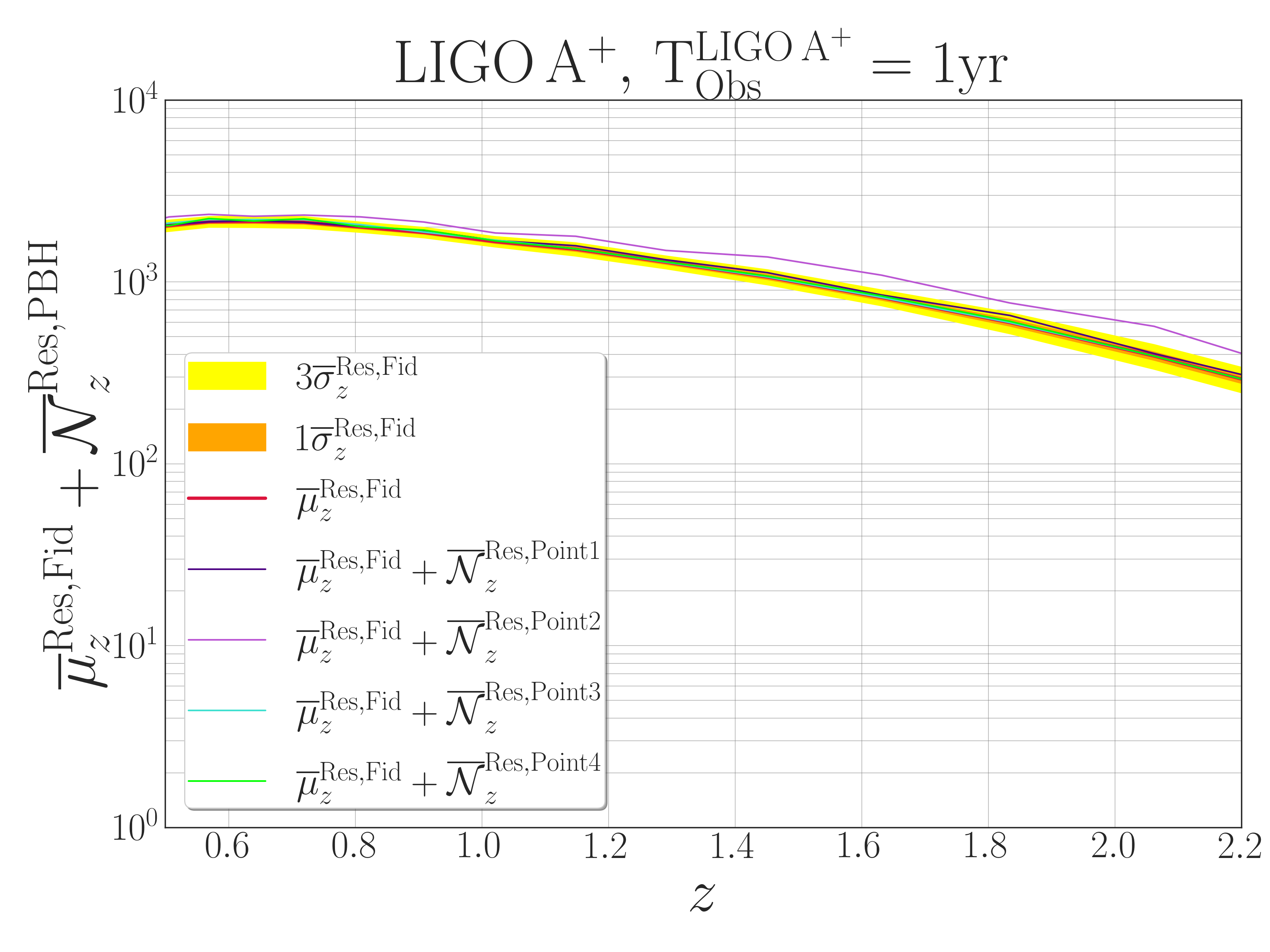} 
   \includegraphics[width=0.46\columnwidth]{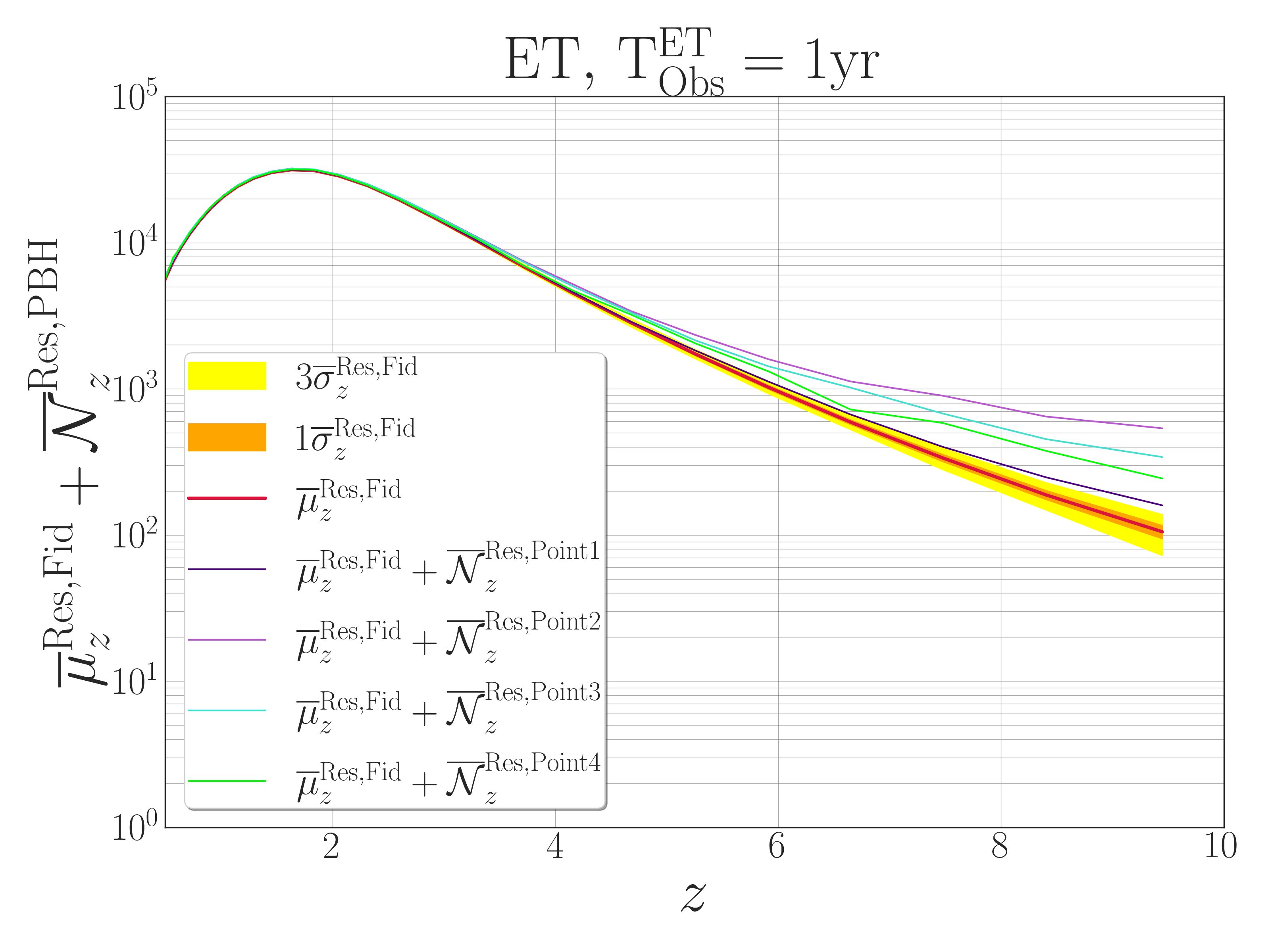}
\caption{\small The impact of the approximations adopted in the semi-analytic method leading to~\Cref{fig:PBHMapsLNLIGO} and~\Cref{fig:PBHMapsLNET}. In the top panels, the quantities are computed with the semi-analytic method.  In the central panels, the quantities are computed on the generated catalogs, but still with the SNR computed with sky and spin-averaging. In the bottom panels, the quantities are computed on the generated catalogs and with the proper SNR evaluation. Each panel also shows the number of events in the fiducial population (red line), with 1 (orange band) and 3$\sigma$ (yellow band) compared with the number of events for the fiducial population plus one of the PBH populations (fixed by the benchmark points). All the results shown in this plot are obtained assuming 1 year of either LIGO A${}^{+}$ (left panels) or ET (right panels) measurements.}
\label{fig:AnNum}
\end{figure}

The results obtained on these benchmark points using LIGO $\rm A^+$ and ET are shown in~\Cref{fig:AnNum}. The three left panels of the figure show the results for LIGO $\rm A^+$, while the right panels  contain the results for ET. The three rows correspond to three different techniques (with increasing levels of accuracy) for the evaluation of the (expected) number of resolvable sources for a given population. The lines in the top row are computed using the semi-analytic method described in~\cref{sec:Meth}, and, in particular, by evaluating~\cref{eq:N_events_binned}. The error bands are estimated from these numbers by assuming a Poisson distribution. Since this procedure is the one used to generate the colored regions in~\Cref{fig:PBHMapsLNLIGO} and~\Cref{fig:PBHMapsLNET},
the population event profiles appearing in these top panels are precisely those establishing the ``minimum $z$" of the benchmarks in those figures (\Cref{tab:BenchPoints} quotes such redshifts).\\

The central panels show the resolvable event distributions determined from the generated catalogs, but still using the  SNR$_{\rm avg}$ approximation. This approximation is the same that we use to get the top-panel results. By comparing the top and central panels, we can assess the impact of realization dependence on our results. As expected, we find consistency in regions with many resolvable sources (small $z$) and deviations in the low-statistics regime (large $z$).\\ 

Finally, the bottom panels show the results obtained using the catalogs and the complete expression for the SNR, consistently including all waveform variables. Deviations between top (equivalent to central) and bottom-panel results manifest at higher redshift. This behavior is expected since small differences in the source parameters can move borderline sources inside or outside the detection threshold. Thus, including all parameters in the SNR evaluation makes results more dependent on the realization effects in the low-statistics regime. The figure shows that this effect is more pronounced for LIGO $\rm A^+$, which has less resolvable sources at high redshift, than for ET. Despite this effect, we observe general good agreement between numerical and semi-analytical results.\\
\begin{figure}[tb]
  \centering
   \includegraphics[width=0.46\columnwidth]{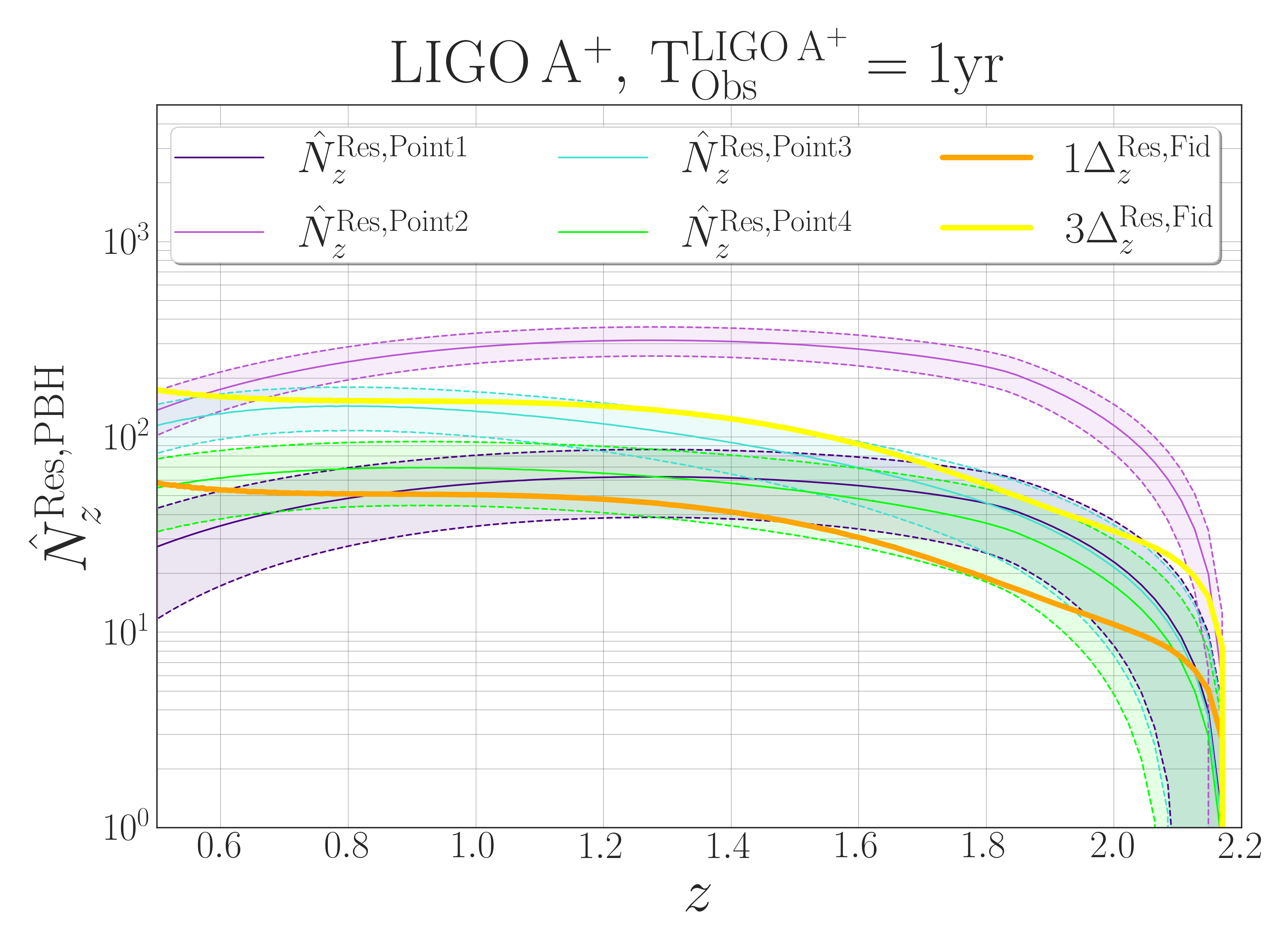} 
   \includegraphics[width=0.46\columnwidth]{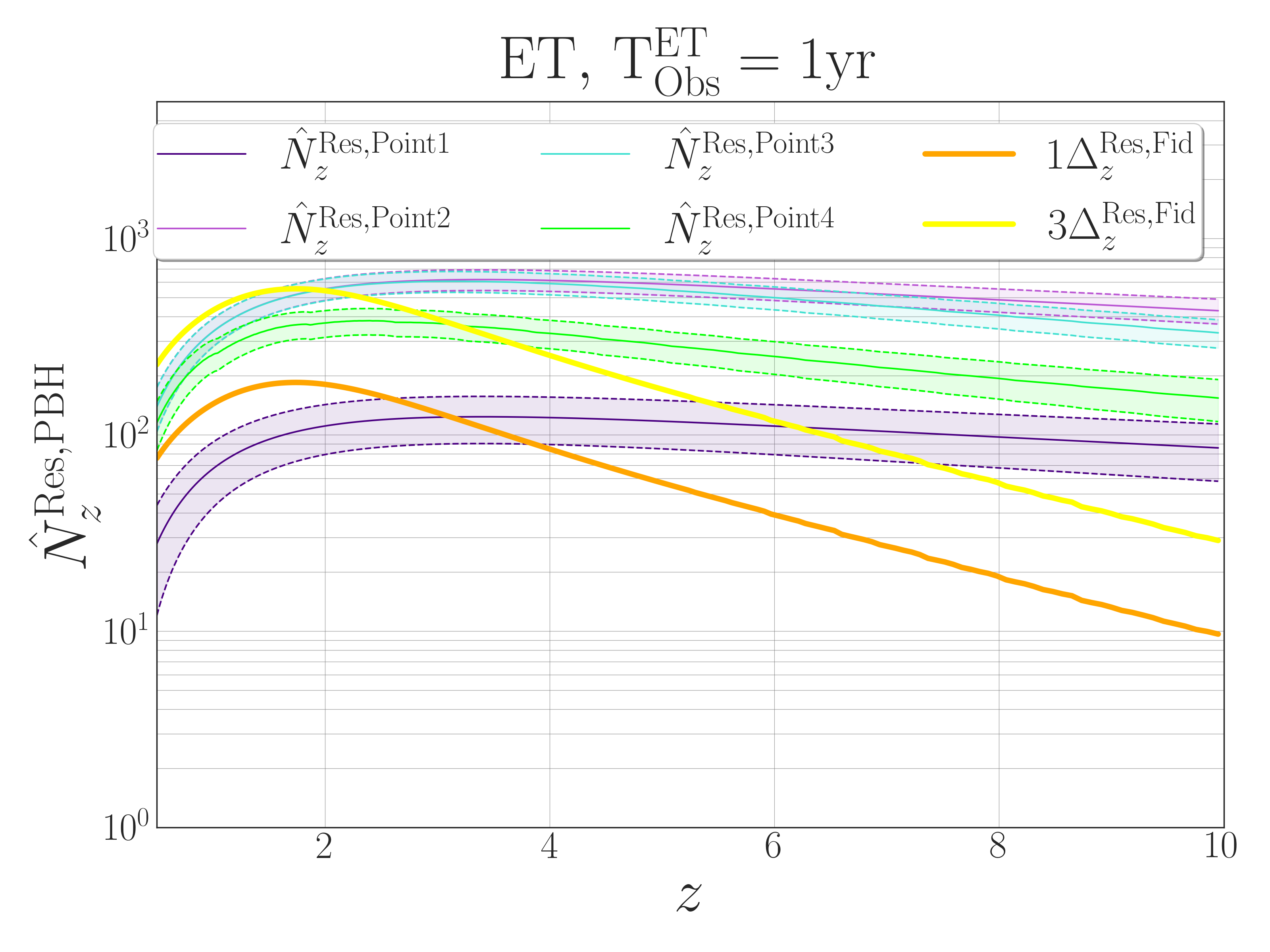}
\caption{Semi-analytical prediction of the number of resolvable events predicted in our benchmark points compared with the analytical estimate for the Poissonian error on the fiducial population. The dashed lines delimit the $3 \Delta^{\rm Res, Point \, i}_z$ region for the i-th benchmark point. The LIGO A${}^{+}$ (ET) results for 1 year of observations are shown in the left (right) panel.}
\label{fig:BenchmarkResDist}
\end{figure}
\begin{figure}[tb]
    \centering
    \includegraphics[width=0.46\columnwidth]{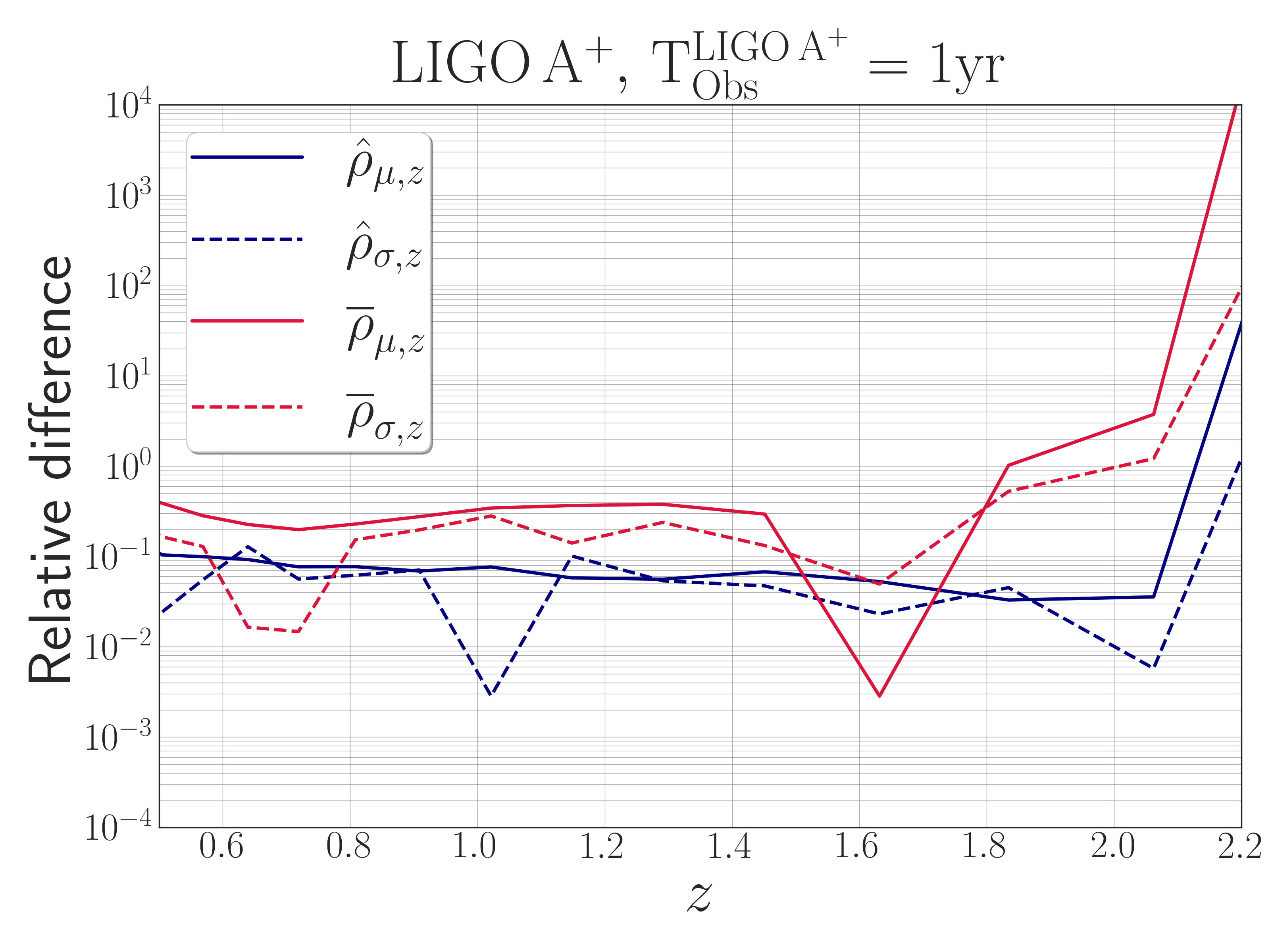}
    \includegraphics[width=0.46\columnwidth]{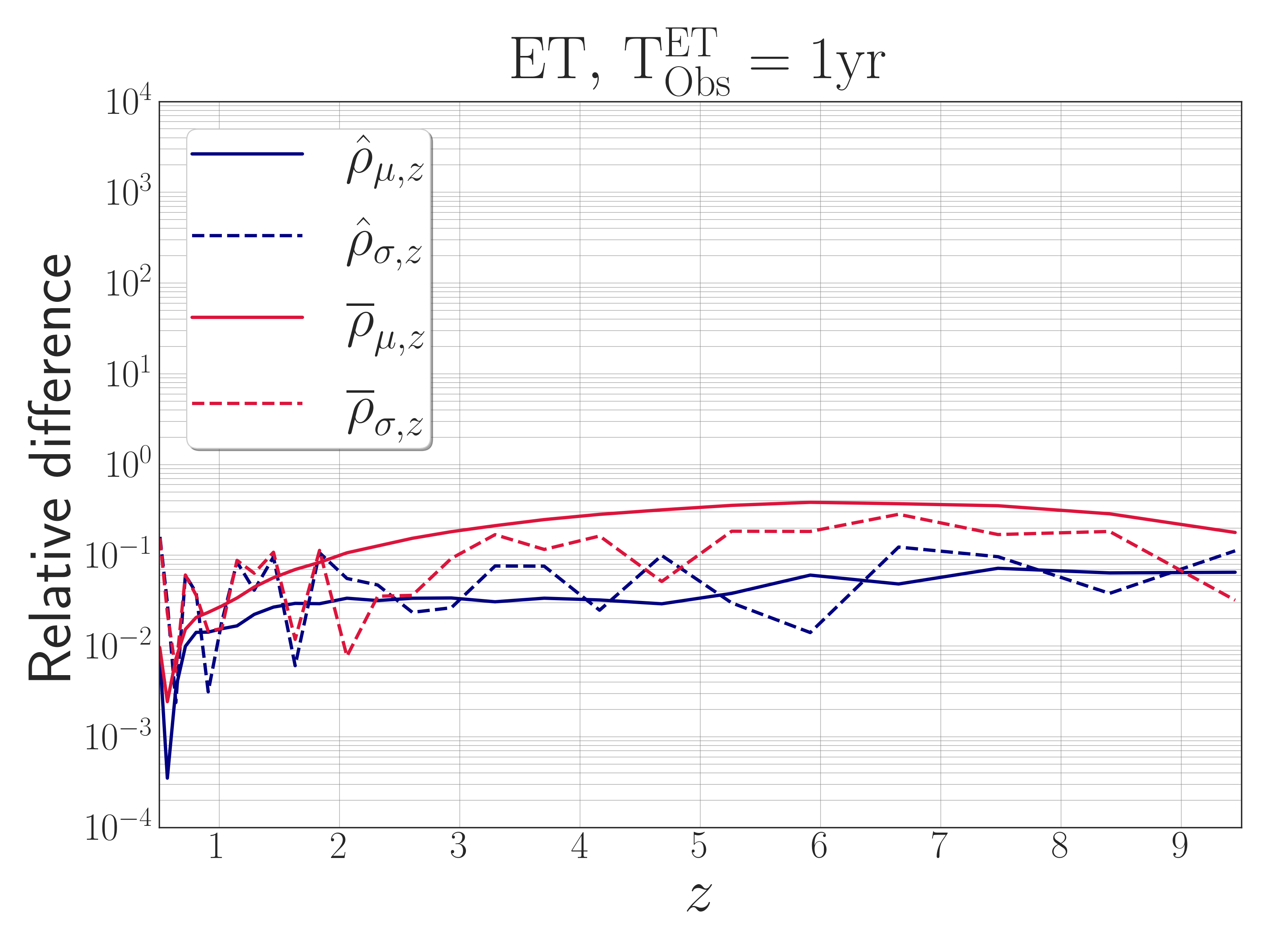}
    \caption{Plots of $\rho_\mu$ (solid lines) and $\rho_\sigma$ (dashed lines) as defined in~\cref{eq:rho_def}, for LIGO A${}^{+}$ (left panel) and ET (right panel). The blue (red) curves use the sky and spin-averaged (full) expression for the SNR.\label{fig:RelativeErrorComparison}}
    
\end{figure}

The actual number of PBH resolvable events per se is a meaningful quantity. \Cref{fig:BenchmarkResDist} shows them. Concretely, it displays their redshift evolutions and their Poissonian uncertainties when these quantities are computed through the semi-analytic method.
It also includes the uncertainty $\Delta^{\rm Res, Fid}_z$ (at 1 and 3 $\sigma$-level) for the fiducial population. The figure gives an insight into how the realization dependencies on the PBH populations would influence the conclusions reached via our semi-analytic analysis. We can see that the impact is non-negligible but does not change our main conclusions.\\

Before concluding, we further scrutinize the relative differences due to the SNR approximation and the realization dependencies. For this purpose, we define the following quantities:
\begin{equation}
\label{eq:rho_def}
  \hat\rho_{\mu, z} = \frac{ |\hat\mu^{\rm Res, Fid}_z - \hat{N}_z^{\rm Res, Fid}|}{\hat\mu_z^{\rm Res, Fid}} \, , \qquad 
  \hat\rho_{\sigma, z} = \frac{|\hat\sigma_z^{\rm Fid, Res} - \sqrt{ \hat{N}_z^{\rm Fid, Res}}| }{\hat\sigma_z^{\rm Fid, Res} }\, ,
\end{equation}
\begin{equation}
\label{eq:rho_def}
  \bar\rho_{\mu, z} = \frac{ |\bar\mu^{\rm Res, Fid}_z - \hat{N}_z^{\rm Res, Fid}|}{\bar\mu_z^{\rm Res, Fid}} \, , \qquad 
  \bar \rho_{\sigma, z} = \frac{|\bar\sigma_z^{\rm Fid, Res} - \sqrt{ \hat{N}_z^{\rm Fid, Res}}| }{\bar\sigma_z^{\rm Fid, Res} }\, .
\end{equation}
\Cref{fig:RelativeErrorComparison} shows their values as a function of redshift when they are computed on the fiducial population.
%We conclude this section by showing, in~\Cref{fig:RelativeErrorComparison}, the relative differences between the semi-analytical estimates of the number of resolvable sources $\hat{N}^{\rm Res,  Fid}$ (and its standard deviation $\sqrt{\hat{N}^{\rm Res,  Fid}}$) and its value estimated using the generated catalogs $\mu^{\rm Res,  Fid}$ and (and its standard deviation $\sigma^{\rm Res,  Fid}$) defined as
%\begin{equation}
%\label{eq:rho_def}
%  \rho_{\mu, z} = \frac{ |\mu^{\rm Res, Fid}_z - \hat{N}_z^{\rm Res, Fid}|}{\mu_z^{\rm Res, Fid}} \, , \qquad \rho_{\sigma, z} = \frac{|\sigma_z^{\rm Fid, Res} - \sqrt{ \hat{N}_z^{\rm Fid, Res}}| }{\sigma_z^{\rm Fid, Res} }\, .
%\end{equation}
%We remind that the quantities $\hat{\mu}^{\rm Res, Fid}$ and $\hat{\sigma}^{\rm Res, Fid}$ are computed assuming the sky and spin-averaged version of the SNR.
The blue curves display ``hat" quantities, and red curves to "bar" quantities.  We see that before reaching the low-statistics regime, the blue curves are always smaller than $10\%$ for both LIGO A${}^{+}$ and ET. This proves a quite good agreement between the semi-analytical approach and the generated catalogs. On the other hand, the red curves 
reach higher values  ($\simeq 30 / 40\%$), so the SNR approximation can induce larger errors than the realization dependence. Nevertheless, both the SNR approximation and the realization dependence are sufficiently small not to jeopardize the main conclusions that one would reach via the semi-analytic method.
%use the value of $\overline{\mu}^{\rm Res, Fid}$ and $\overline{\sigma}^{\rm Res, Fid}$ evaluated using the full expression for the SNR. While we notice some level of degradation with respect to the blue curves (the relative difference is up to $\simeq 30 / 40\%$), we still find quite good agreement between the results.

\section{Conclusions}
\label{sec:Conclusions}
In this paper, we have discussed the prospects of detecting potential PBHB populations with future GW detectors. For this purpose, we have assumed a fiducial population in agreement with the GWTC-3 results and added PBHB populations with different merger rates and mass distributions to test whether these would lead to observable signatures in LIGO $\rm A^+$, ET, or LISA. Using Earth-based detectors, we have checked how the number of resolvable events changes in the presence of PBHB populations. In particular, we have evaluated this analytically and tested our results by simulating event catalogs to assess the impact of low statistics on the analytic results. We have generally found good agreement between our semi-analytical and numerical results. Beyond that, we have evaluated the increase in the amplitude of the SGWB arising from the PBH contribution and tested its detectability with LISA using a FIM approach. We found that the information LISA will bring might be significant to test whether the SGWB is due to SOBHBs only.\\

For all models considered in this work, we have found sizable regions of the parameter spaces where the PBHB populations will lead to significant variations in the number of detectable events in LIGO $\rm A^+$ and ET (with ET performing better in most cases) with respect to SOBHB expectations. However, it is worth stressing that detecting events at high redshift does not imply that it will always be possible to infer their redshift accurately~\cite{Mancarella:2023ehn}. Moreover, we have found that for all models considered in this work, there are sizable parts of the parameter spaces leading to an increase in the SGWB amplitude that would be detectable with LISA. Interestingly, since the SGWB integrates over all masses, the SGWB measurement can also test populations with very low and very large masses, generating signals beyond the reach of future Earth-based detectors. Indeed, different GW detectors probe complementarily distinct parts of the parameter space. In particular, our results highlight three different regimes:
\begin{itemize}
\item \textbf{\textrm{Signatures in Earth-based detectors and LISA:} } This can happen if the PBH population becomes abundant (but still statistically marginal in present LVK observations) around (or after) the SFR peak so that the number of individual events does not decline at $z\simeq 2$. Simultaneously, the SGWB at LISA exceeds the SOBHB prediction.

\item \textbf{Signatures in Earth-based detectors only:} The PBH population grows very slowly with $z$ and becomes sizable only at high redshifts. In this case, the signals from unresolved sources are faint, and their contribution to the SGWB at LISA is not sufficiently strong to modify the SOBHB prediction significantly. While deviations in the merger rate might be appreciated, their statistical significance might not be sufficient to pin down the presence of a secondary population unless some additional features are found in, \emph{e.g.}, the mass distribution of the observed population. 

\item \textbf{Signatures in LISA only:} The PBH population has very small or very large masses, and the signals are not detectable with Earth-based detectors. LISA observes an SGWB amplitude incompatible with the value predicted using the SOBHB population measured by Earth-based detectors.

\end{itemize}

Overall, we conclude that the considered measurements from Earth-based detectors and LISA will generally be complementary, and our understanding of the BH population that we observe in our Universe will improve if we use their synergy. This conclusion comes with no surprise, knowing the underlying differences between the properties of the signals these detectors will probe. Furthermore, we generally observe that the dependency of the LISA SGWB on the population parameters scales differently than the distribution of resolvable sources that Earth-based interferometers will detect. This fact implies that, in general, SGWB measurements will help Earth-based detectors improve the constraints on the BH population parameters that we observe in our Universe.\\

As discussed in~\cref{sec:Meth}, our analysis does not include errors on the fiducial population parameter, which are currently quite broad on some of the most influencing parameters and would impact our results significantly if extrapolated to the 
%observable 
volume that LIGO $\rm A^+$ and ET detector will probe. However, the open codes presented in our GitHub repository~\cite{GithubCodesPaolo} can be readily updated when the new results of future inference papers, \emph{e.g.}, by the LVK collaboration, come out. With improvements in the LVK network, we expect more (and more accurate) detections, which will reduce the uncertainties on the population parameters, making the analysis much more reliable. \\

The present study assumes that the two populations do not interact with one another, \emph{i.e.}, mergers only involve BHs drawn from the same population. This assumption impacts both the number and the properties of the mergers. By dropping this assumption and assuming the two populations to have sufficiently different mass ranges, it would be possible to enhance the number of Extreme Mass Ratio Inspirals (EMRIs) \footnote{For SGWB predictions and cosmological constraints with EMRIs, see, \emph{e.g.},~\cite{Babak:2017tow, Chen:2021ool, Laghi:2021pqk, Pozzoli:2023kxy,  Liu:2023onj}.
} in the LISA band significantly~\cite{Gair:2017ynp, Babak:2017tow, Guo:2022sdd, Mazzolari:2022cho}. Thus, determining the abundance of these objects can also be used to further constrain the eventual presence of PBHB populations on GW detectors with LISA-like frequency range. Moreover, keeping track of the number of resolvable sources in different frequency bands (\emph{e.g.}, BHs with masses higher than the ones considered in the present work would merge in the LISA frequency band) could also provide a possible tracer for the presence of PBHB population. \\

We conclude by commenting on alternative methods to assess the detectability of PBHB populations beyond the ones considered in this work for both Earth-based and Space-based detectors. It would be possible, in principle, to adopt other criteria similar to the one we have introduced. For example, since we expect PBHBs to become relevant at high redshift, variations on the cumulative number of resolvable sources predicted after a given redshift could provide a viable alternative. While maintaining less information on the source distribution, this approach would be less sensitive to the error in the inference of the source distance~\cite{Mancarella:2023ehn}. Finally, a fully consistent population analysis based on hierarchical Bayesian modeling would provide a robust and more accurate alternative to the criteria discussed here. Thus, we deem it worth exploring these (and possibly other) methodologies in future works.

\acknowledgments
We thank Gabriele Franciolini for several useful comments on a draft of this work.
MP acknowledges the hospitality of Imperial College London, which provided office space during some
parts of this project. PM acknowledges the Cost action CA16104 for financing the STSM to London that led to the start of this work. GN is partly supported by
the grant Project.~No.~302640 funded by the Research Council of Norway.

\appendix

\section{The SOBHB fiducial population}
\label{app:FidMod}

In this appendix, we detail the fiducial SOBHB population model we adopt throughout the analysis. In most aspects, we follow the approach of ref.~\cite{Babak:2023lro} and rely on the master equation in~\cref{eq:GeneralSourceGenerator}. We proceed by clarifying the functional forms of the quantities appearing in such an equation.\\

As already discussed in the main text, the current LVK data put tight bounds on the SMBHB population properties up to $z \sim 0.5$ \cite{LIGOScientific:2021djp}. Few events have been detected at redshift $0.5 \lesssim z \lesssim 1$, but they are too rare and/or poorly reconstructed to impose strong constraints~\cite{LIGOScientific:2020iuh, OBrien:2021sua}. Despite these caveats, current data are compatible with a population of SOBHBs with a merger rate behaving as
\begin{equation}
    R(z) = R_0 (1 + z)^\kappa
\end{equation} 
for $z\lesssim 0.5$, with $R(z = 0.2) = 28.3^{+13.9}_{- 9.1} {\rm Gpc}^{-3} {\rm  yr}^{-1}$ and $\kappa = 2.9^{+ 1.7}_{- 1.8}$~\cite{Abbott:2020gyp, LIGOScientific:2021djp}. At higher redshift, $R(z)$ has to keep track of the stellar-formation origin of the binaries and, to some degree, resemble the SFR. Thus, consistently with~\cite{Babak:2023lro}, we choose the Madau-Dickinson phenomenological profile~\cite{Madau:2014bja, Mangiagli:2019sxg} with a negligible time delay\footnote{Other choices~\cite{Dominik:2012kk, neijssel2019effect, Dvorkin:2016wac, Mapelli:2017hqk, Santoliquido:2020bry, vanSon:2021zpk, Fishbach:2021mhp} are possible and might qualitatively change the results, but not the rationale, of our analysis.}. Such a choice leads to
\begin{equation}
    R_{\rm SOBHB}(z) = R_{0} \frac{(1 + z)^\kappa}{1 + ((1+z)/2.9)^{\kappa + 2.9}}\,,
\label{eq:MergingRate}
\end{equation}   
where $R_{0}$ is set so that $R(z=0.2)$ matches the measured value.\\

For what concerns the mass distribution, we adopt the \emph{power law + peak} scenario~\cite{Abbott:2020gyp, LIGOScientific:2021djp}
\begin{equation}
\begin{aligned}
p(m_1,m_2| & m_{m}, m_{M}, \alpha, \beta_q, \mu_m, \sigma_m, \delta_{m}, \lambda_{\rm peak})  =\\
& C_{\rm mass} \, \pi_1(m_1| \alpha, \mu_m, \sigma_m, m_{m}, m_{M}, \delta_{m}, \lambda_{\rm peak})
\,  \pi_2(q|\beta_q, m_1, m_{m}, \delta_{m})\,,
\label{eq:MassDistribution}
\end{aligned}
\end{equation}  
with $C_{\rm mass}$ being a normalization constant and $q$ being the mass ratio $q = m_2/m_1$. The functions $\pi_1$ and $\pi_2$ read as
\begin{equation}
\begin{aligned}
    \pi_1(m_1| \alpha, \mu_m, \sigma_m, m_{m}, m_{M}, \delta_{m}, \lambda_{\rm peak}) & =  \\ 
    \left[ (1 - \lambda_{\rm peak})\mathfrak{P}(m1|-\alpha, m_{M}) \right. & \left.+\lambda_{\rm peak}\mathfrak{G}(m_1|\mu_m,\sigma_m)\right] \mathfrak{S}(m1|m_{m}, \delta_{m})\,
\label{eq:MassDistribution1}
\end{aligned}
\end{equation}  
and

\begin{equation}
\pi_2(q|\beta_q, m_1, m_{m}, \delta_{m}) = C_{q}(m_1) \, q^{\beta_q} \, \mathfrak{S}(m_2|m_{m}, \delta_{m})\,,
\label{eq:MassDistribution2}
\end{equation}  
where $C_{q}(m_1)$ is a normalization function, $\mathfrak{S}$ is a smoothing function for the low mass cutoff, and $\mathfrak{P}$ and $\mathfrak{G}$ are respectively a normalized power law and a normalized Gaussian distribution 
\begin{equation}
\mathfrak{P} = C_{\rm PL}m^{- \alpha}\,,
\label{eq:MassPL}
\end{equation}  
\begin{equation}
\mathfrak{G} = \frac{C_{m}}{ \sqrt{2 \pi \sigma_{m}^2 } }\, \exp \left[ -\frac{1}{2} \left(\frac{m - \mu_{m}}{\sigma_{m}} \right)^2 \right] \,,
\label{eq:MassGS}
\end{equation}  
with $\alpha$ being the spectral index of the power law, $\mu_{m}$ and $\sigma_{m}$ being the mean and width of the Gaussian, and
$C_{\rm PL}$ and $C_{m}$ being normalization.
The smoothing function $\mathfrak{S}$ imposes a smooth cutoff for low masses, rising from $0$ to $1$ in the interval $[m_{m}, m_{m} + \delta_{m}]$
\begin{equation}
\mathfrak{S} =\begin{cases} 0, & \mbox{if } m < m_{m} \\ 
[f(m - m_{m}, \delta_{m}) + 1]^{-1}, & \mbox{if } m \in [m_{m}, m_{m} + \delta_{m}] \\
1, & \mbox{if } m > m_{m} + \delta_{m} 
\end{cases}\,,
\label{eq:MassSmooth}
\end{equation}  
with
\begin{equation}
f(m^\prime, \delta_{m}) = \exp{\left( \frac{\delta_{m}}{m^\prime} + \frac{\delta_{m}}
{m^\prime - \delta_{m}} \right)} \, ,
\label{eq:MassSmooth2}
\end{equation}  
so that, by construction, we have $m \ge m_m$. The high end of the mass range doesn't have an explicit cutoff, but large masses are statistically suppressed. In practice, we set $m_{M}=100 M_\odot$, which is slightly higher than the values used in the LVK analysis~\cite{KAGRA:2021duu}, to take into account possible higher mass events of astrophysical origin.\footnote{We test that other choices would not practically change our results. For \emph{e.g.},~$m_{M}=150 M_\odot$, no masses above 100 $M_\odot$ appear in our catalog realizations due to the PDF suppression.}
All the hyperparameters entering the mass distribution~\cref{eq:MassDistribution} are fixed at the central values of the LVK analysis outcome reported in~\cref{tab:MassPDFParms}.\\
\begin{table}
\centering
\begin{tabular}{|c|c|c|c|c|c|c|c|c|}
\hline
Parameter & $m_{m} \, [M_{\odot}] $ & $m_{M} \, [M_{\odot}]$ &  $\delta_{m} \, [M_{\odot}]$ & $\lambda_{\rm peak}$ & $\alpha$ & $\beta_q$ & $\mu_m$ & $\sigma_m$ \\ \hline
Value & $ 5.0^{+0.86}_{-1.7} $ & 
 $ 100 $ & 
$ 4.9^{+3.4}_{-3.2} $ & 
 $ 0.038^{+ 0.058}_{-0.026} $ & 
$ 3.5^{+0.6}_{-0.56} $ & 
 $ 1.1^{+1.7}_{-1.3}$ & 
 $ 34^{+2.6}_{-4.0} $ & 
$ 5.69^{+4.28}_{-4.34}$ \\ \hline
\end{tabular}
\caption{Fiducial values (with 1$\sigma$ C.L.) for the mass function hyperparameters~\cite{KAGRA:2021duu}.}
\label{tab:MassPDFParms}
\end{table}

\begin{table}
\centering
\begin{tabular}{|c|c|c|c|c|c|c|}
\hline
Coefficients & $a_{M} $ & E[$a$] & Var[$a$] & $\zeta $ & $\sigma_1$ & $\sigma_2$ \\ \hline
 Value & $ 1$ & 
 $ 0.26^{+0.09}_{-0.07} $ &
 $ 0.02^{+0.02}_{-0.01} $ &
 $ 0.76^{+0.22}_{-0.45} $ & 
 $ 0.87^{+1.08}_{-0.45} $ & 
 $ 0.87^{+1.08}_{-0.45} $ \\ \hline

\end{tabular}
\caption{Fiducial values (with 1$\sigma$ C.L.) for the spin amplitude and spin tilt hyperparameters~\cite{KAGRA:2021duu}.}
\label{tab:spin_params}
\end{table}

The spin distribution is a product of two different PDFs, one for the spin amplitudes and one for the spin tilts. 
The former reads as~\cite{Abbott:2020gyp, LIGOScientific:2021djp}:
\begin{equation}
    p(a_i|\alpha_a,\beta_a) = \frac{a_i^{\alpha_a -1}(1-a_i^{\beta_a - 1})}{B(\alpha_a,\beta_a)}\,,
\label{eq:SpinDistribution}
\end{equation}
where $B(\alpha_a,\beta_a)$ is a Beta function that guarantees the appropriate normalization of the PDF. The $\alpha_a$ and $\beta_a$ are positive constants defined through
\begin{equation}
    \begin{array}{lcl} $E$[a] & = & \frac{\alpha_a}{\alpha_a + \beta_a} \\ $Var$[a] & = & \frac{\alpha_a \beta_a}{(\alpha_a + \beta_a)^2 (\alpha_a + \beta_a +1)} \end{array}\,,
\label{eq:SpinParameters}
\end{equation}
where $E[a] $ and $Var[a] $ are set in~\cref{tab:spin_params}. We stress that the spin amplitudes of the two black holes are independent of one another. On the other hand, the PDF spin tilt distribution reads
\begin{equation}
    p(\cos(t_1),\cos(t_2)|\sigma_1,\sigma_2,\zeta) = \frac{1 - \zeta}{4} + \frac{2 \zeta}{\pi} \prod\limits_{i \in {1,2}} \frac{ \exp\left\{  -\left[1 - {\rm cos} (t_i) \right]^2 / (2 \sigma_i^2) \right\} }{\sigma_i \, { \rm erf}(\sqrt{2} /\sigma_i)}\,,
\label{eq:SpinTiltDistribution}
\end{equation}
which is a mixture between an isotropic and a truncated Gaussian distribution centered in $\cos(t_i) \approx 1$. The $\sigma_i$, and $\zeta$ parameters are specified in~\cref{tab:spin_params}.\\

Finally, for cosmology we assume the $\Lambda$CDM model where the Hubble parameter is $H_0 = h \times 100 \, {\rm km/(s \,Mpc) }$, with $h = 0.678$ being its dimensionless value, and $\Omega_m = 0.3$ and $\Omega_\Lambda = 0.7$.

\section{PBH contribution to the Dark Matter relic abundance}
\label{app:ConvMaps}
\begin{figure}[th!]
  \centering
   \begin{subfigure}[b]{\textwidth}
   \includegraphics[width=\columnwidth]{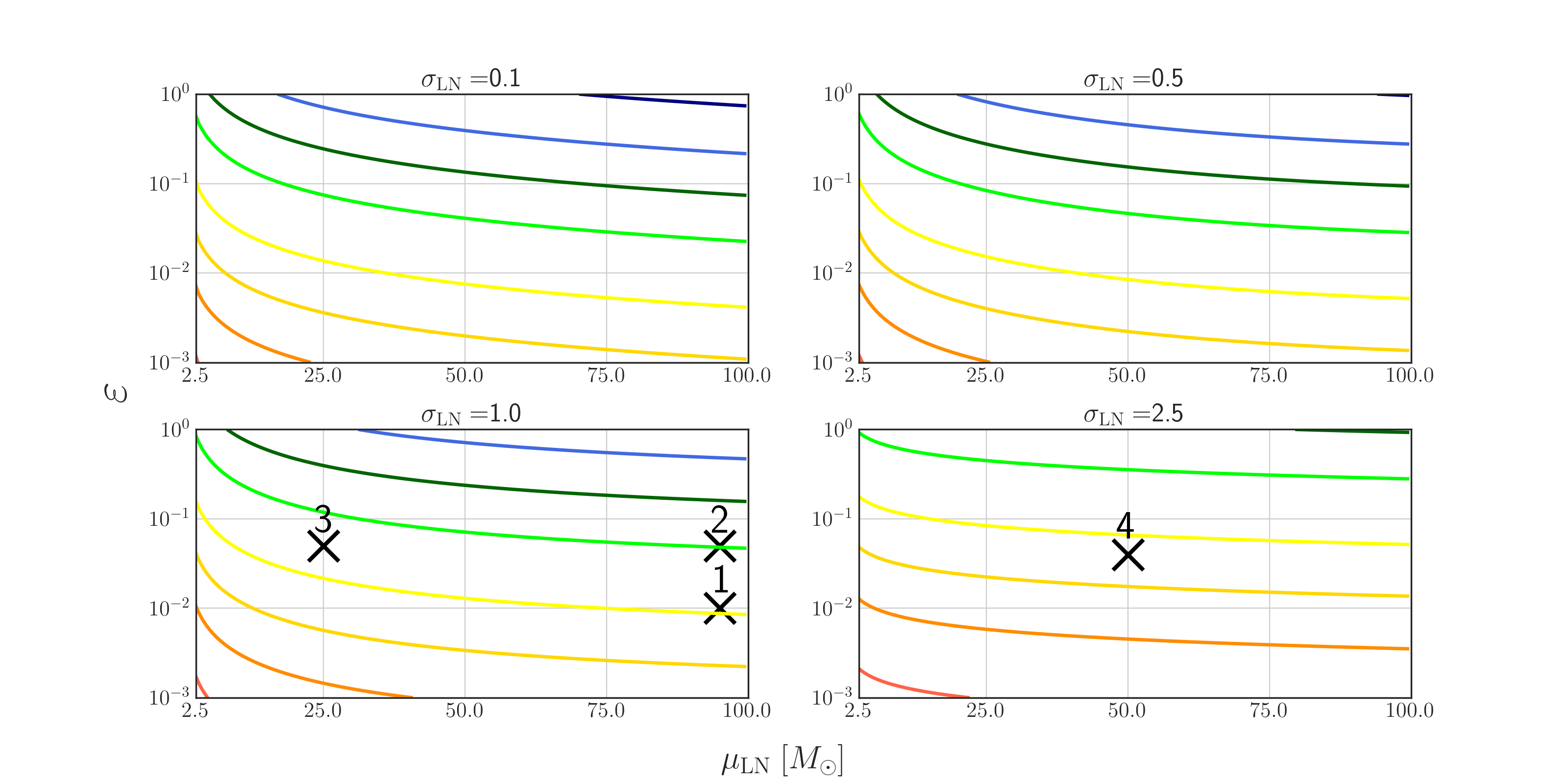}
   \end{subfigure}
   \begin{subfigure}[b]{\textwidth}
   \centering
   \includegraphics[width=0.75\columnwidth]{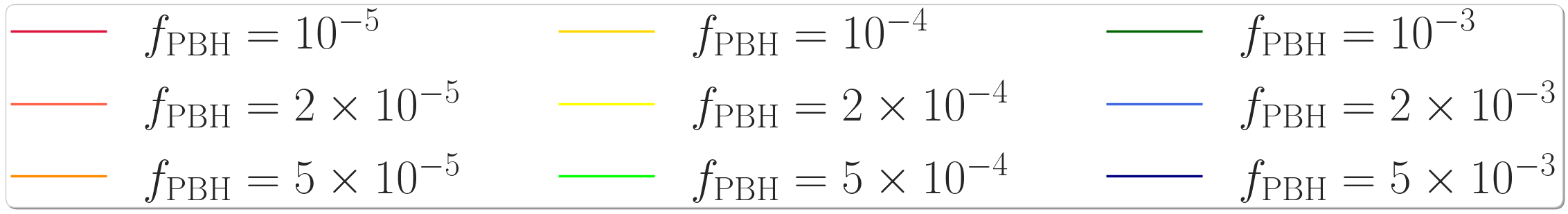}
   \end{subfigure}
\caption{Conversion maps from $\varepsilon$ to $f_{\rm PBH}$ of the parameter spaces of PBHB populations with LN mass function. Each panel corresponds to a different value of $\sigma_{\rm LN}$, and it spans over values of $\varepsilon$ and $\mu_{\rm LN}$. Crosses indicate the benchmark points used in the analysis. \label{fig:PBHConversionMaps}}
\end{figure}

While PBHs behave as cold DM and could, at least in principle constitute a sizable amount of the presently observed DM, their abundance in the stellar mass range is tightly constrained~\cite{Nakamura:1997sm, Ioka:1998nz, EROS-2:2006ryy, Ricotti_2008, Wyrzykowski_2011, Ali-Haimoud:2017rtz, Garcia-Bellido_2017, Villanueva-Domingo:2021spv}. We define $f_{\rm PBH} \equiv  \Omega_{\rm PBH} / \Omega_{\rm DM}$, the ratio between today's PBH and DM energy densities in the Universe. For a given PBH population model, the parameters $\varepsilon$ and $f_{\rm PBH}$ can be explicitly related. We obtain their relationship in the case that our PBH population, which we derive from phenomenological models,  is dominated by binaries that gravitationally decoupled from the
Hubble flow before the matter–radiation equality~\cite{Hutsi:2020sol, Franciolini:2023opt, LISACosmologyWorkingGroup:2023njw}. \\

In~\cref{eq:PBH_Rtz}, we modulate the PBHB merger rate $R_{\rm PBHB}(z)$ through the parameter $\varepsilon$. An alternative way to write $R_{\rm PBHB}(z)$ is~\cite{Hutsi:2020sol, Franciolini:2023opt} 
%
%\begin{equation}
%\begin{split}
%    R^{Full}_{\rm PBHB}(z, f_{\rm PBH}) = & \frac{1.6 \times 10^6}{\rm Gpc^3 yr} f_{\rm PBH}^{53/37} \left[ \frac{t(z)}{t(z = 0)}\right]^{-34 / 37} \int \textrm{d} m_1 \int \textrm{d} m_2 \left[ \left(\frac{m_1 + m_2}{M_\odot}\right)^{-32 / 37} \times \right.\\
 %   & \times \left. \eta^{-34/37} \, S(m_1, m_2, z, f_{\rm PBH}, \Phi_{\rm LN/G}) \, \Phi_{\rm LN/G}(m_1) \,\Phi_{\rm LN/G}(m_2) \right ] \, .
    \label{eq:full_Rpbh}
%\end{split}
%\end{equation}
%
\begin{equation}
\begin{split}
    R_{\rm PBHB}(z) = & \frac{1.6 \times 10^6}{\rm Gpc^3\, yr} f_{\rm PBH}^{53/37} \left[ \frac{t(z)}{t(z = 0)}\right]^{-\frac{34}{37}} \int \textrm{d} m_1 \int \textrm{d} m_2 \left[ \left(\frac{m_1 + m_2}{M_\odot}\right)^{-\frac{32}{37}}  \eta^{-\frac{34}{37}} \mathcal{S} \right],
    \label{eq:full_Rpbh}
\end{split}
\end{equation}
where $\eta = m_1 m_2 /(m_1 + m_2)^2$ and
$\mathcal{S} = \Phi_{\rm LN}(m_1) \,\Phi_{\rm LN}(m_2) S$.
The function $\Phi_{\rm LN}$ is given in \cref{eq:PBH_LNMassPDF}. The function $S$ is a suppression factor accounting for environmental effects that slow down the binary formation or favor their disruption. It can be approximated as\footnote{In general, in $\mathcal{S}$ contains an extra suppression factor, which introduces redshift dependence at small $z$. Such a term is negligible for $f_{\rm PBHB}\simeq 10^{-3}$~\cite{Hutsi:2020sol, Franciolini:2023opt}, which is the region of parameter space relevant parts for most models considered in this work. As a consequence, we neglect this factor.}
\begin{equation}
    S_1 \approx 1.42 \left( \frac{\left< m^2 \right>/\left< m \right>^2}{\Bar{N}(m_1, m_2, f_{\rm PBH}) + C} + \frac{\sigma^2_{\rm M}}{f_{\rm PBH}^2}\right)
    e^{-\Bar{T}}\, ,
    \label{eq:PBHSupp1}
\end{equation}
with
\begin{equation}
    \Bar{T} = \frac{m_1 + m_2}{\left< m \right>} \left( \frac{f_{\rm PBH}}{f_{\rm PBH} + \sigma_{\rm M}}\right)\, ,
    \label{eq:PBHSupp1NBar}
\end{equation}
\begin{equation}
    C = \frac{\left< m^2 \right>}{\sigma^2_{\rm M} \left< m \right>^2} \frac{f_{\rm PBH}^2}{\left[ \frac{\Gamma(29/37)}{\sqrt{\pi}}U\left(\frac{21}{74}, \frac{1}{2}, \frac{5 f_{\rm PBH}^2}{6 \sigma_{\rm M}^2}\right)\right]^{-74/21}-1}\,.
    \label{eq:PBHSupp1C}
\end{equation}
Here $\sigma_{\rm M}\approx 0.004 $ represents the rescaled variance of matter density perturbations at the
time the binaries form, $\Gamma$ denotes the Euler gamma function, $U(a, b, z)$ is the confluent hypergeometric function, while $\left< m \right>$ and $\left< m^2 \right>$ are the first and second momenta of the PBH mass PDF.\\ 

The comparison of \cref{eq:PBH_Rtz} to \cref{eq:full_Rpbh} yields
\begin{equation}
 \varepsilon = \frac{1.6 \times 10^6}{R_0} f_{\rm PBH}^{53/37} \int \textrm{d} m_1 \int \textrm{d} m_2 \left[ \left(\frac{m_1 + m_2}{M_\odot}\right)^{-\frac{32}{37}}  \eta^{-\frac{34}{37}} \mathcal{S} \right].
\end{equation}
\Cref{fig:PBHConversionMaps} shows the values of $f_{\rm PBH}$ that arise in the parameter regions of the PBH models considered in our analysis.

\section{Detector characteristics}
\label{sec:detectors}
In this study, we consider LIGO A${}^+$ and ET as representatives of upcoming and future Earth-based interferometers and LISA as a reference for the first generation of space-based GW detectors. The precise timeline and operational durations of these instruments are uncertain. Nevertheless, it is reasonable to anticipate that LIGO A${}^+$ will operate for several years before the early/mid-2030s when ET and LISA are expected to commence to acquire data. In the lack of a well-defined progress plan, we consider a couple of somewhat extreme timeline scenarios, believing that the actual future will likely fall somewhere in between. Concretely, we analyze 1 and 10 years of data for LIGO A${}^+$ and ET, and 4 and 10 for LISA. We leave it to the knowledge of the future reader to estimate which scenario the future will tend to and in which order each detector and its measurements will arrive.\\

\paragraph{Earth-based interferomenters}
\begin{figure}[t!]
  \centering
  \includegraphics[width=\columnwidth]{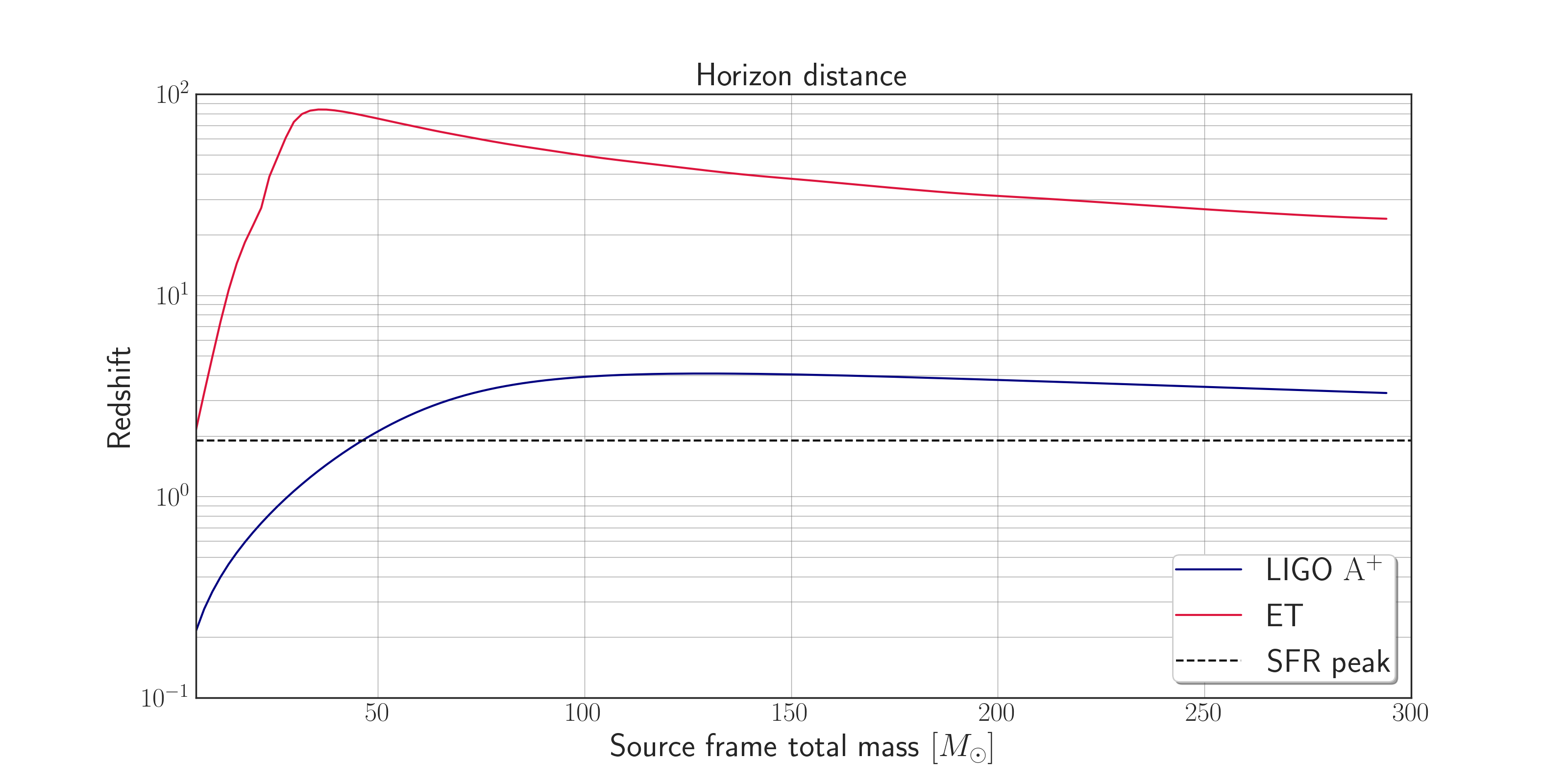}
\caption{Horizon distance for the LIGO A${}^{+}$ (blue) and ET (red) detectors as a function of the total
mass arising in the populations of our analysis. The black dashed line corresponds to the redshift
of the peak of the SFR (with no time delays) used in our fiducial SOBHB population. \label{fig:horizon_distance}}
\end{figure}

The location of the two LIGO A${}^{+}$ detectors are set to be in the Livingston (N 30${}^\circ$330${}^\prime$, W 90${}^\circ$460${}^\prime$) and  Hanford (N 46${}^\circ$270${}^\prime$, W 119${}^\circ$240${}^\prime$) sites. Regarding the sensitivity, we use the curve described in ref.~\cite{LIGOSensCurve}, with frequency range $[5, 5000]$\,Hz. For ET, we assume the location proposed in the Sos Enattos mine in the Lula area (N 40${}^\circ$260${}^\prime$, E 9${}^\circ$260${}^\prime$) with the ET-D-sum 
sensitivity in the frequency range $[0.1, 10^4]$\,Hz~\cite{Hild_2011, ETSensCurve}. However, our resolvable event analysis is nearly independent of the precise detector sites. The expected horizon distance for these detector configurations, w.r.t the BHB populations considered in this paper, is presented in fig.~\ref{fig:horizon_distance}\\

\paragraph{LISA}
LISA will be the first interferometer in space. The detector will consist of three satellites orbiting around the Lagrange point L5. For our analysis, we assume mission adoption sensitivity in the frequency range $[3 \times 10^{-5}, 0.5]$\,Hz~\cite{LISAsrd}. In the following, we describe the $2$-parameters instrument noise model~\cite{LISA:2017pwj, Babak:2021mhe} based on the results of the LISA Pathfinder mission, as well as the latest laboratory test. In particular, we report the LISA sensitivity in the Time Delay Interferometry (TDI) channels A and T\footnote{TDI is a technique designed for LISA to suppress the otherwise dominant (and several orders of magnitude larger than the required noise levels) primary noises. TDI consists of combining interferometric measurements performed at different times. It can be shown that for a fully symmetric LISA configuration, it is possible to introduce an orthogonal (\emph{i.e.}, noise in the different channels is uncorrelated) TDI basis, typically dubbed AET. See, \emph{e.g.}, refs.~\cite{Armstrong_1999, Tinto:1999yr, Estabrook:2000ef, Tinto:2020fcc, Prince:2002hp, Shaddock:2003dj, Tinto:2003vj} for details.}.
For more details on the noise model and the TDI construction, see \emph{e.g.}, ref.~\cite{Flauger:2020qyi, Hartwig:2023pft}.\\

The noise in LISA is a combination of two main components: Test Mass (TM) acceleration noise and Optical Metrology System (OMS) noise. The power spectra $P_{\rm TM}$, $P_{\rm OMS}$, for these two components are
\begin{equation}
\begin{aligned}
P_{\rm TM}(f, A) & = A^2 \frac{ {\rm fm^2} }{s^4 \textrm{Hz}} \left[1 + \left( \frac{0.4 \textrm{mHz}}{f} \right)^2 \right]
\left[1 + \left( \frac{f}{8\textrm{mHz}} \right)^4 \right] \left( \frac{1}{2 \pi f} \right)^4
\left( \frac{2 \pi f}{c} \right)^2 \,, \\
P_{\rm OMS}(f, P) & = P^2 \frac{\rm pm^2}{\textrm{Hz}} \left[1 + \left( \frac{2 \textrm{mHz}}{f} \right)^4 \right]\left( \frac{2 \pi f}{c} \right)^2 \,,
\label{eq:Noise_spectra}
\end{aligned}
\end{equation}  
where $c$ is the light speed and the two noise parameters $A$ and $P$ control the amplitudes of the TM and OMS components, respectively. The total noise spectral densities in the TDI A and T channels read
\begin{align}
    \label{eq:FshMtrx_Naa}
N_{\rm AA}(f, A, P) & =  8 \sin^2 \left( \frac{2 \pi f L}{c} \right) \bigg \{ 4 \left[ 1 + \cos\left(\frac{2 \pi f L}{c} \right) + \cos^2 \left( \frac{2 \pi f L}{c} \right) \right] P_{\rm TM}(f, A) + \\
\nonumber
& \qquad + \left[ 2 + \cos \left( \frac{2 \pi f L}{c} \right) \right] P_{\rm OMS}(f, P) \bigg \} \,,\\
\label{eq:FshMtrx_Ntt}
N_{\rm TT}(f, A, P) & =  16 \sin^2 \left( \frac{2 \pi f L}{c} \right) \bigg \{ 2 \left[ 1 - \cos\left(\frac{2 \pi f L}{c} \right) \right]^2 P_{\rm TM}(f, A) + \\
\nonumber
 & \qquad + \left[ 1 - \cos \left( \frac{2 \pi f L}{c} \right) \right] P_{\rm OMS}(f, P) \bigg \} \,,
\end{align} 
with $L = 2.5 \, \times 10^{9}$m is the LISA armlength.\\

Given the noise power spectra, the strain sensitivity (for a generical channel $ij$) is defined as
\begin{equation}
S_{n,ij}(f, A, P) = \frac{N_{ij}(f, A, P)}{R_{ij}(f)} = \frac{N_{ij}(f, A, P)}{16 \sin^2 \left( \frac{2 \pi f L}{c} \right) \left( \frac{2 \pi f L}{c} \right)^2 \tilde{R}_{ij}(f)} \,,
\label{eq:FshMtrx_SNoise}
\end{equation}  
where $R_{ij}(f)$ is the (quadratic) response function, mapping incoming GW signals onto the TDI data stream. The response can be further expanded as a purely geometrical factor, $\tilde{R}_{ij}(f)$, times TDI-dependent terms. While $\tilde{R}_{ij}(f)$ should be evaluated, approximate expressions for the A and T channels read
\begin{equation}
\tilde{R}_{\rm AA}(f) = \frac{9}{20} \frac{1}{1 + 0.7 \left( \frac{2 \pi f L}{c} \right)^2} \,, \qquad \qquad
\tilde{R}_{\rm TT}(f) = \frac{9}{20} \frac{\left( \frac{2 \pi f L}{c} \right)^6}{1.8 \times 10^3 + 0.7 \left( \frac{2 \pi f L}{c} \right)^8} \,.
\label{eq:FshMtrx}
\end{equation} 
Since $\tilde{R}_{\rm TT}$ is strongly suppressed at low frequencies with respect to $\tilde{R}_{\rm AA}$, the T channel is typically assumed to be signal insensitive. 
It is customary to express the noise in $\Omega$ units using
\begin{equation}
h^2 \tilde{\Omega}_{n, ij} (f, A, P) = \frac{4 \pi^2 f^3}{3 (H_0 / h)^2} S_{n, ij} (f, A, P) \, .
\label{eq:FshMtrx_Omegan}
\end{equation}  
In the analyses presented in this work, we assume the fiducial values for the noise parameters to be $A = 3$, $P=15$ with 20$\%$ Gaussian priors.

\section{Analytical derivation of the SGWB from a population of merging objects}
\label{app:analytical_SGWB}

In this appendix, we summarize the analytical derivation of the SGWB sourced by the PBHB and SOBHB populations. Further details on the derivation can be found in refs.~\cite{Phinney:2001di, Sesana:2008mz} for details. We focus on the case of the LISA detector, for which we know that the number of resolvable sources is small enough ($\sim 10$ sources for $\rm SNR_{\rm tresh}= 8$) to prevent dampenings of the SGWB signal in the high-frequency region of the LISA sensitivity band~\cite{Babak:2023lro}.\\

Following the approach of refs.~\cite{1989thyg.book.....H, Phinney:2001di}, the characteristic spectral strain of the SGWB can be defined as
\begin{equation}
 h^2_c(f) = \frac{4G}{\pi c^2 f^2} \int_{0}^{\infty} dz \frac{\textrm{d} n}{\textrm{d} z} \frac{1}{1 + z} f_r \frac{\textrm{d} E_{\rm GW}}{\textrm{d} f_r}\bigg|_{f_r = f(1 + z)} \,.
\label{eq:SGWB_CharStrain}
\end{equation}  
Here $\textrm{d} n/\textrm{d} z$ and $\textrm{d} E_{\rm GW}/\textrm{d} f_r$ are, respectively, the comoving number density of the sources, and the (redshifted) energy per log frequency interval produced by each source. The latter, in the circular-orbit approximation, reads as
\begin{equation}
 \frac{\textrm{d} E_{\rm GW}}{\textrm{d} f_r} = \frac{\pi}{3G} \frac{(G\mathcal{M})^{5/3}}{(\pi f_r)^{1/3}} \bigg|_{f_r = f(1 + z)} \,,
\label{eq:SGWB_dEdf}
\end{equation}  
with $\mathcal{M}$ being the chirp mass of the source. The former is a function of the statistical properties of the source population. It can be
written as
\begin{equation}
\frac{\textrm{d} n}{\textrm{d} z} = 
\int_{0}^{\infty} \textrm{d}\mathcal{M} \frac{\textrm{d}^2 n}{\textrm{d}z\, \textrm{d}\mathcal{M}} = 
\int_{0}^{\infty} \textrm{d}\mathcal{M} \,R(z) \, P[\mathcal{M}(m_1, m_2)]\frac{\textrm{d} t_r}{\textrm{d} z} \,,
\label{eq:SGWB_dndz}
\end{equation}  
where $R(z)$ is the merger rate of the population, $P[\mathcal{M}(m_1, m_2)]$ is the probability of having a chirp mass $\mathcal{M}$ from the $m_1$ and $m_2$ mass PDFs, and $\textrm{d} t_r/ \textrm{d} z$ is a drift term defined as
\begin{equation}
\frac{\textrm{d} t_r}{ \textrm{d} z} = \frac{1}{H_0(1 + z)\sqrt{\Omega_m(1 + z)^3 + \Omega_\Lambda}} \,.
\label{eq:SGWB_dtrdz}
\end{equation}

By putting all these elements together, one finds
\begin{equation}
h^2_c(f) = \frac{4 G^{5/3}}{3 c^2 \pi^{1/3} f^{4/3}} \int_{0}^{\infty} \textrm{d} z \int_{m_{m}}^{m_{M}} \textrm{d} m_1 \int_{m_{m}}^{m_{M}} \textrm{d} m_2 R(z) p(m_1, m_2) \frac{M(m_1,m_2)^{5/3}}{(1 + z)^{1/3}} \frac{\textrm{d} t_r}{\textrm{d} z} \,.
\label{eq:SGWB_hcFinal}
\end{equation}  
This can be rephrased in $\Omega$ units as 
\begin{equation}
\Omega_{\rm GW}(f) = \frac{2 \pi^2  f^2 \, h_c^2(f)}{3 H_0^2} \,.
\label{eq:SGWB_OmegaUnits}
\end{equation}  

\bibliographystyle{JHEP}
\bibliography{references.bib}

\end{document}